\documentclass[preprint]{aastex}
\usepackage{epsfig}

\begin{document}

\title{A Multi-wavelength View of the TeV Blazar Markarian 421: Correlated 
Variability, Flaring, and Spectral Evolution} 

\author{M.~B{\l}a\.zejowski\altaffilmark{1},
G.~Blaylock\altaffilmark{2},
I.~H.~Bond\altaffilmark{3},
S.~M.~Bradbury\altaffilmark{3},
J.~H.~Buckley\altaffilmark{4},
D.~A.~Carter-Lewis\altaffilmark{5},
O.~Celik\altaffilmark{6},
P.~Cogan\altaffilmark{7},
W.~Cui\altaffilmark{1},
M.~Daniel\altaffilmark{7},
C.~Duke\altaffilmark{8},
A.~Falcone\altaffilmark{1},
D.~J.~Fegan\altaffilmark{7},
S.~J.~Fegan\altaffilmark{6},
J.~P.~Finley\altaffilmark{1},
L.~Fortson\altaffilmark{9},
S.~Gammell\altaffilmark{7},
K.~Gibbs\altaffilmark{10},
G.~G.~Gillanders\altaffilmark{11},
J.~Grube\altaffilmark{3},
K. Gutierrez\altaffilmark{4},
J.~Hall\altaffilmark{12},
D.~Hanna\altaffilmark{13},
J.~Holder\altaffilmark{3},
D.~Horan\altaffilmark{10},
B.~Humensky\altaffilmark{14},
G.~Kenny\altaffilmark{11},
M.~Kertzman\altaffilmark{15},
D.~Kieda\altaffilmark{12},
J.~Kildea\altaffilmark{13},
J.~Knapp\altaffilmark{3},
K.~Kosack\altaffilmark{4},
H.~Krawczynski\altaffilmark{4},
F.~Krennrich\altaffilmark{5},
M.~Lang\altaffilmark{11},
S.~LeBohec\altaffilmark{12},
E.~Linton\altaffilmark{14},
J.~Lloyd-Evans\altaffilmark{3},
G.~Maier\altaffilmark{3},
D.~Mendoza\altaffilmark{1},
A.~Milovanovic\altaffilmark{3},
P.~Moriarty\altaffilmark{16},
T.~N.~Nagai\altaffilmark{12},
R.~A.~Ong\altaffilmark{6},
B. Power-Mooney\altaffilmark{7},
J.~Quinn\altaffilmark{7},
M.~Quinn\altaffilmark{16},
K.~Ragan\altaffilmark{13},
P.~T.~Reynolds\altaffilmark{17},
P.~Rebillot\altaffilmark{4},
H.~J.~Rose\altaffilmark{3},
M.~Schroedter\altaffilmark{5},
G.~H.~Sembroski\altaffilmark{1},
S.~P.~Swordy\altaffilmark{14},
A.~Syson\altaffilmark{3},
L.~Valcarel\altaffilmark{13},
V.~V.~Vassiliev\altaffilmark{6},
S.~P.~Wakely\altaffilmark{14},
G.~Walker\altaffilmark{12},
T.~C.~Weekes\altaffilmark{10}, 
R.~White\altaffilmark{3}, and
J.~Zweerink\altaffilmark{6},
\\and \\
B.~Mochejska\altaffilmark{1}, B.~Smith\altaffilmark{1},
M.~Aller\altaffilmark{18}, 
H.~Aller\altaffilmark{18}, H.~Ter\"{a}sranta\altaffilmark{19},
P.~Boltwood\altaffilmark{20}, A.~Sadun\altaffilmark{21}, 
K.~Stanek\altaffilmark{22}, E.~Adams\altaffilmark{22}, 
J.~Foster\altaffilmark{22}, J.~Hartman\altaffilmark{22}, 
K.~Lai\altaffilmark{22}, M.~B\"{o}ttcher\altaffilmark{23}, 
A.~Reimer\altaffilmark{24}, and I.~Jung\altaffilmark{4}
}
\altaffiltext{1}{Department of Physics, Purdue University, West Lafayette,  
IN 47907, USA}
\altaffiltext{2}{Department of Physics, University of Massachusetts, 
Amherst, MA 01003, USA}
\altaffiltext{3}{Department of Physics, University of Leeds, Leeds, LS2 9JT,
Yorkshire, England, UK}
\altaffiltext{4}{Department of Physics, Washington University, St.~Louis, 
MO 63130, USA}
\altaffiltext{5}{Department of Physics and Astronomy, Iowa State University,
Ames, IA 50011, USA}
\altaffiltext{6}{Department of Physics, University of California, Los 
Angeles, CA 90095, USA}
\altaffiltext{7}{Physics Department, National University of Ireland,
Belfield, Dublin 4, Ireland}
\altaffiltext{8}{Physics Department, Grinnell College, Grinnell, IA 50112, USA}
\altaffiltext{9}{Adler Planetarium and Astronomy Museum, Astronomy Department,
Chicago, IL 60605}
\altaffiltext{10}{Fred Lawrence Whipple Observatory, Harvard-Smithsonian
Center for Astrophysics, Amado, AZ 85645, USA}
\altaffiltext{11}{Physics Department, National University of Ireland, Galway,
Ireland}
\altaffiltext{12}{High Energy Astrophysics Institute, University of Utah, 
Salt Lake City, UT 84112, USA}
\altaffiltext{13}{Physics Department, McGill University, Montreal, QC H3A 2T8,
Canada}
\altaffiltext{14}{Enrico Fermi Institute, University of Chicago, Chicago,
IL 60637, USA}
\altaffiltext{15}{Physics Department, DePauw University, Greencastle, 
IN, 46135, USA}
\altaffiltext{16}{School of Physical Sciences, Galway-Mayo Institute of 
Technology, Galway, Ireland}
\altaffiltext{17}{Department of Applied Physics and Instrumentation, Cork 
Institute of Technology, Cork, Ireland}
\altaffiltext{18}{Department of Astronomy, University of Michigan, Ann Arbor, 
MI 48109, USA}
\altaffiltext{19}{Mets\"ahovi Radio Observatory, Helsinki University of 
Technology, Mes\"ahovintie 114, 02540 Kylm\"al\"a, Finland}
\altaffiltext{20}{Boltwood Observatory, 1655 Main Street, Stittsville, 
Ont. K2S 1N6, Canada}
\altaffiltext{21}{Department of Physics, University of Colorado at Denver, 
CO 80217, USA}
\altaffiltext{22}{Harvard-Smithsonian Center for Astrophysics, 60 Garden 
Street, Cambridge, MA 02138, USA}
\altaffiltext{23}{Department of Astronomy, Ohio University, Athens, OH 05701,
USA}
\altaffiltext{24}{Institut f\"ur Theoretische Physik, Lehrstuhl IV: Weltraum 
und Astrophysik, Ruhr-Universit\"at Bochum, D 44780 Bochum, Germany}

\begin{abstract}
\noindent
We report results from an intensive multi-wavelength monitoring campaign
on the TeV blazar Mrk 421 over the period of 2003--2004. The source was 
observed simultaneously at TeV energies with the Whipple 10 m telescope 
and at X-ray energies with {\em Rossi X-ray Timing Explorer} (RXTE) 
during each clear night within the {\em Whipple} observing windows. Supporting 
observations were also frequently carried out at optical and radio 
wavelengths to provide simultaneous or contemporaneous coverages. The 
large amount of simultaneous data has allowed us to examine the variability 
of Mrk 421 in detail, including cross-band correlation and broad-band 
spectral variability, over a wide range of flux. The variabilities are 
generally correlated between the X-ray and gamma-ray bands, although the 
correlation appears to be fairly loose. The light curves show the presence 
of flares with varying amplitudes on a wide range of timescales both at 
X-ray and TeV energies. Of particular interest is the presence of TeV 
flares that have no coincident counterparts at longer wavelengths, because 
the phenomenon seems difficult to understand in the context of the proposed 
emission models for TeV blazars. We have also found that the TeV flux 
reached its peak days {\em before} the X-ray flux did during a giant flare
(or outburst) in 2004 (with the peak flux reaching $\sim$135 mCrab in X-rays, 
as seen by the ASM/RXTE, and $\sim$3 Crab in gamma rays). Such a difference 
in the development of the flare presents a further challenge to the leptonic 
and hadronic emission models alike. Mrk 421 varied much less at optical and 
radio wavelengths. Surprisingly, the normalized variability amplitude in 
optical seems to be comparable to that in radio, perhaps suggesting the 
presence of different populations of emitting electrons in the jet. The 
spectral energy distribution of Mrk 421 is seen to vary with flux, with 
the two characteristic peaks moving toward higher energies at higher fluxes. 
We have failed to fit the measured SEDs with a one-zone SSC model; introducing 
additional zones greatly improves the fits. We have derived constraints on 
the physical properties of the X-ray/gamma-ray flaring regions from the 
observed variability (and SED) of the source. The implications of the results 
are discussed.
\end{abstract}


\section{Introduction}

Over the past decade or so, one of the most exciting advances in high energy
astrophysics has been the detection of sources at TeV energies with
ground-based gamma ray facilities (see Weekes 2003 for a recent review). 
Among the sources detected, blazars are arguably the most intriguing. They 
represent the only type of active galactic nuclei (AGN) that has been 
detected at TeV energies (although a 4-$\sigma$ detection of M87 has been
reported; Aharonian et al. 2003). To date, there are a total of six firmly 
established TeV blazars.

The emission from a blazar is generally thought to be dominated by radiation 
from a relativistic jet that is directed roughly along the line of sight 
(review by Urry \& Padovani 1995 and references therein). Relativistic 
beaming is necessary to keep gamma-ray photons from being significantly 
attenuated by the surrounding radiation field (via photon-photon pair 
production). The spectral energy distribution (SED) of TeV blazars invariably 
shows two characteristic peaks in the $\nu F_{\nu}$ representation, with one 
located at X-ray energies and the other at TeV energies (Fossati et al. 1998).
There seems to be a general correlation between the two SED peaks as the 
source varies (e.g., Buckley et al. 1996; Catanese et al. 1997; Maraschi et 
al. 1999; Petry et al. 2000). 

A popular class of models associates the X-ray emission from a TeV blazar 
with synchrotron radiation from highly relativistic electrons in the jet and 
the TeV emission with inverse-Compton scattering of the synchrotron photons 
by the electrons themselves (i.e., synchrotron self-Compton or SSC for short; 
Marscher \& Gear 1985; Maraschi et al. 1992; Dermer et al. 1992; Sikora et 
al. 1994; see B\"{o}ttcher 2002 for a recent review). The SSC models can, 
therefore, naturally account for the observed X-ray--TeV correlation. 
Moreover, they have also enjoyed some success in reproducing the measured 
SEDs. However, the models still face challenges in explaining some of the 
observed phenomena, such as the presence of ``orphan'' TeV flares 
(Krawczynski et al. 2004; Cui et al. 2004). 

Alternatively, the jet might be energetically dominated by the magnetic 
field and it is the synchrotron radiation from highly relativistic protons 
that might be responsible for the observed TeV gamma rays (Aharonian 2000; 
M\"ucke et al. 2003). Other hadronic processes have also been considered, 
including photo-meson production, neutral pion decay, and synchrotron-pair 
cascading (e.g., Mannheim \& Biermann 1992; M\"ucke et al. 2003), but they 
are thought to be less important in TeV blazars (Aharonian 2000; M\"ucke et 
al. 2003). Another class of hadronic models invokes $pp$ processes, for 
instance, in the collision between the jet and ambient ``clouds'' 
(e.g., Dar \& Laor 1997; Beall \& Bednarek 1999) or inside the (dense) jet 
(Pohl \& Schlickeiser 2000). In this case, the gamma-ray emission is mainly
attributed to the decay of neutral pions produced in the $pp$ interactions. 
In both classes of hadronic models, the emission at 
X-ray and longer wavelengths is still attributed to the synchrotron 
radiation from relativistic electrons (and positrons) in the jet, as in the 
SSC models. Although the hadronic models may also be able to describe the 
observed SED of TeV blazars and accommodate the X-ray--TeV correlation, 
they are generally challenged by the most rapid gamma-ray variabilities 
observed in TeV blazars (Gaidos et al. 1996). 

TeV blazars are also known to undergo flaring episodes both at X-ray and 
TeV energies. The flares have been observed over a wide range of timescales, 
from months down to less than an hour. The observed X-ray flaring hierarchy 
seems to imply a scale-invariant physical origin of the flares (Cui 2004; 
Xue \& Cui 2005). Blazar flares are thought to be related to internal shocks 
in the jet (Rees 1978; Spada et al. 2001), or to the ejection of 
relativistic plasma into the jet (e.g., B\"ottcher et al. 1997; Mastichiadis 
\& Kirk 1997). Recently, it is suggested that the flares could also be 
associated with magnetic reconnection events in a magnetically dominated 
jet (Lyutikov 2003) and thus they could be similar to solar flares in this 
regard. Such a model might offer a natural explanation for the hierarchical 
flaring phenomenon, again in analogy to solar flares. 

To make further progress on distinguishing the emission models proposed 
for TeV blazars, we believe that a large amount of simultaneous or 
contemporaneous data is critically needed over a wide range of flux, 
especially in the crucial X-ray and TeV bands, for quantifying the SED 
and spectral variability of a 
source and for allowing investigations of such important issues as 
variability timescales, cross-band correlation, spectral variability,  
spectral hysteresis, etc. Such data are severely lacking at present, despite 
intense observational efforts over the years. In this paper, we present
results from an intensive multi-wavelength monitoring campaign on Mrk 421. 
This source is the first TeV blazar discovered (Punch et al. 1992) and 
remains one of the few blazars that can be detected at TeV energies nearly 
all the time with ground-based imaging atmospheric Cherenkov telescopes 
(IACTs). Some of the preliminary results have appeared elsewhere (Cui et al. 
2004); they are superseded by those presented in this work. 

We have assumed the following values for the various cosmological parameters: 
$H=71\mbox{ }km\mbox{ }s^{-1}\mbox{ }Mpc^{-1}$, $\Omega_m=0.27$, and 
$\Omega_{\Lambda}=0.73$. The corresponding luminosity distance of Mrk 421 
($z=0.031$) is about 129.8 Mpc.

\section{Observations and Data Reduction}

\subsection{Gamma-ray Observations}

From 2003 February to 2004 June, Mrk 421 was observed at TeV energies with 
the Whipple 10 m Telescope (on Mt. Hopkins, AZ) during each clear night 
within the dark moon observing periods. The typical exposure time of a 
nightly observation was 28 minutes, corresponding to one observing run, 
but more runs were taken on occasion, especially 
near the end of the campaign (in 2004 April) when the source was seen to 
undergo an usually large X-ray outburst as seen by the All-Sky Monitor (ASM) 
on {\em RXTE}. To achieve simultaneous coverages of the source both in the 
TeV and X-ray bands, we communicated with the {\em RXTE Science Operations 
Facility} to ensure that the {\em Whipple} and {\em RXTE} observing 
schedules were 
matched as closely as possible. A total of 306 runs were collected in good 
(empirically designated as ``A'' or ``B'') weather, and roughly 80\% of 
them were taken at zenith angles $\lesssim$30\arcdeg.

The procedure for reducing and analyzing {\em Whipple} data has been 
standardized 
over the years (Hillas 1985; Reynolds et al. 1993; Mohanty et al. 1998).
A detailed description of the current hardware can be found in Finley et al. 
(2001). For clarity, we briefly summarize a few key points that are relevant 
to this work. The success of IACTs lies in the fact that the Cherenkov 
images of an air shower produced by a gamma-ray primary have different 
shapes and orientations than those found in an air shower produced by 
cosmic-ray particles (mostly protons).  The images in gamma-ray events are 
typically more compact and are more aligned to point towards the position 
of the source than those in cosmic-ray events. In practice, the image shape 
of an event is characterized by the major and minor axes of the best-fit 
ellipse to the image. The orientation of the ellipse is 
characterized by the parameter "$\alpha$", defined as the angle between the 
major axis and the line connecting the center of the ellipse to the center 
of the field-of-view (FOV). We have developed standard selection criteria
based primarily on the image shape and orientation parameters that remove 
over 99\% of the cosmic ray events while keeping about half of the gamma-ray 
events (see, e.g., Falcone et al. 2004 for more details).

The {\em Whipple} observations are conducted in one of the two modes: 
tracking and 
ON/OFF. In the tracking mode, the telescope tracks the target across the sky 
so that the source stays at the center of the FOV throughout the observation. 
In the ON/OFF mode, on the other hand, the telescope tracks the target only 
during the ON run; it is then offset by 30 minutes in right ascension
during the OFF run, and tracks the field as it covers the same range of 
zenith and azimuthal angles. The OFF run provides a direct measurement of 
the background, although the difference in the sky brightness of the fields 
between the ON and OFF runs must be taken into account by using a technique 
known as software padding (Reynolds et al 1993). For tracking 
observations, the background is derived from the $\alpha$ histogram of the 
events that have passed all but the orientation cuts. Since real gamma-ray 
events from a source should all be concentrated at small $\alpha$ values 
($<$ 15\arcdeg) for on-axis observations, the background level can be 
estimated from events with larger $\alpha$ values (20\arcdeg--65\arcdeg), if 
the ratio of the number of background events in the two ranges of the 
$\alpha$ parameter is known. This ratio is derived from observations of 
fields in which there is no evidence for a gamma-ray source (see, e.g., 
Falcone et al. 2004).

Nearly all of the {\em Whipple} observations reported in this work were 
carried 
out in the tracking mode. We followed the standard procedure just described 
to obtain count rates from individual runs (taken in the good weather) and 
thus the long-term light curves. To examine variability on short timescales, 
we sometimes sub-divided runs into time intervals shorter than the nominal
28 minutes and constructed light curves with correspondingly smaller time 
bins. To correct for the effects due to changes in the zenith angle and 
overall throughput of the telescope, we applied the method developed by 
LeBohec \& Holder (2003) to 
the data. We should note that for a given season the corrections were made 
with respect to a (somewhat arbitrarily chosen) reference run taken at 
30\arcdeg\ zenith angle during a clear night. To correct for changes in 
the telescope throughput across seasons, we used the measured rates for the 
Crab Nebula to further calibrate the light curves. For reference, the rate 
of the Crab Nebula is about 2.40 and 2.93 $\gamma\mbox{ }min^{-1}$ for 
the 2002/2003 and 2003/2004 seasons, respectively. 

The spectral analysis was carried out by following Method 1 described in 
Mohanty et al. (1998). The technical aspects and difficulties involved 
in finding matching pairs for the spectral analysis of tracking observations 
are explained in detail by Petry et al. (2002) and Daniel et al. (2005). 
Combined with the fact that the observations were taken only in the tracking 
mode, poor statistics made it extremely challenging to derive a TeV spectrum 
from observations taken at low fluxes. For those cases, the tolerance for 
the parameter cuts was tightened to be just 1.5 standard deviations from the 
average value of the simulations, as opposed to the usual 2 standard 
deviations, in order to further reduce the cosmic-ray background (Daniel et 
al. 2005). The tighter cuts still retained $\sim 80\%$ of gamma rays. The 
downside is that the effective collection area of the telescope is not as 
independent of energy as can be ideally hoped for (Mohanty et al. 1998).

\subsection{X-ray observations}

In coordination with each {\em Whipple} observation, we took a snapshot of 
Mrk 421 at X-ray energies with {\em RXTE} (with a typical exposure time 
of 2--3 ks). We should note, however, that not every X-ray observation 
was accompanied by a simultaneous {\em Whipple} observation (due, e.g., to 
poor weather) and vice versa (especially near the end of the campaign,
when many more {\em Whipple} observations were made to monitor the source in 
an exceptionally bright state; see Fig.~1). For this work, we only used 
data from the PCA instrument on {\em RXTE}, which covers a nominal energy 
range of 2--60 keV. The PCA consists of five nearly identical 
proportional counter units (PCUs). However, only two of the PCUs, PCU 0 
and PCU 2, were in use throughout our campaign, due to operational 
constraints. PCU 0 has lost its front veto layer, so the data from it are 
more prone to contamination by events caused by low-energy electrons 
entering the detector. The problem is particularly relevant to variability 
studies of relatively weak sources, such as Mrk~421. For this work, 
therefore, we have chosen PCU 2 as our ``standard'' detector for flux 
normalization and spectral analysis.

We followed Cui (2004) closely in reducing and analyzing the PCA data. 
Briefly, the data were reduced with {\em FTOOLS 5.2}. For a given
observation, we first filtered data by following the standard procedure
for faint sources,\footnote{See the online {\em RXTE} Cook Book, available 
at http://heasarc.gsfc.nasa.gov/docs/xte/recipes/cook\_book.html.} which
resulted in a list of good time intervals (GTIs). We then simulated
background events for the observation by using the latest background
model that is appropriate for faint sources. Using the GTIs, we proceeded 
to extract a light curve for each PCU separately. We repeated the steps to 
construct the corresponding background light curves from the simulated 
events. We then subtracted the background from the total to obtain the 
light curves of the source. Following a similar procedure, for each
observation, we also constructed the X-ray spectrum for each PCU and its 
associated background spectrum. In this case, however, we only used 
data from the first xenon layer of each PCU (which is most accurately 
calibrated), which limits the spectral coverage to roughly 2.5--25 keV. 
Since few counts were detected at higher energies, the impact of the 
reduced spectral coverage is very minimal. 

\subsection{Optical observations}

The optical data were obtained with the Fred Lawrence Whipple 
Observatory (FLWO) 1.2 m telescope (located adjacent to the Whipple 10 m 
gamma-ray telescope on Mt. Hopkins) and with the 0.4 m telescope at the 
Boltwood Observatory in Stittsville, Ontario, Canada. We note that we 
had no optical coverage of the source during the 2002/2003 {\em Whipple}
observing season. 

The FLWO 1.2 m was equipped with 4Shooter CCD mosaic with four 
$2048 \times 2048$ chips. Each chip covers a $11\farcm 4\times 11\farcm 4$ 
FOV.  The data were collected during 31 nights, from 2003 December 14 to 
2004 February 17. A total of 77 images were obtained in the $B$ band, 69 
in the $V$ band, 67 in the $R$ band, and 78 in the $I$ band, with an 
exposure time of 30 s for each image. The preliminary processing of the 
CCD frames was performed with the standard routines in the IRAF ccdproc 
package.\footnote{IRAF is distributed by the National Optical Astronomy 
Observatories, which are operated by the Association of Universities for 
Research in Astronomy, Inc., under cooperative agreement with the NSF.} 

Photometry was extracted using the {\sc DAOphot/Allstar} package (Stetson 
1987). The fitting radius for profile photometry was varied with the seeing. 
The median seeing in our dataset was 4.5\arcsec. A detailed description of 
the applied procedure is given in \S~3.2 of Mochejska et al.\ (2002). The 
derivation of photometry was complicated by the proximity of two very 
bright stars, HD~95934 (V=6 mag) and HD~95976 (V=7.5 mag), at angles of 
2.1\arcmin\ and 4\arcmin\ from Mrk 421, respectively. These angles 
correspond to distances of 187 and 356 pixels on the images. The presence 
of these stars introduces a gradient in the 
background, approximately in the north-south direction. To estimate the 
magnitude of this gradient, we examined two $10 \times 10$ pixel regions 
located at distances of 30--40 pixels north and south of Mrk 421. These 
regions were chosen to coincide with the annulus of 18-45 pixels used by 
{\sc Allstar} for background determination. The difference in counts 
between the two regions is at a level of 1.4\%, 0.7\%, 0.3\%, and 0.2\% 
of the peak value of Mrk 421 with BVRI filters, respectively, on our 
best seeing images, and of 3.9\%, 1.9\%, 1.1\%, and 0.6\% on images with 
seeing close to the upper $85^{th}$ percentile. Thus, it varies with the 
seeing in BVRI by 2.5\%, 1.2\%, 0.8\%, and 0.4\%, respectively, all very 
small compared with the variability amplitudes of the source.
The transformations of instrumental magnitudes to the standard system
were derived from observations of 27 stars in 3 Landolt (1992) standard
fields, collected on 19 January 2004. The systematic errors associated 
with the calibration are estimated to be 0.02 for BVI and 0.01 for R.

The Boltwood 0.4 m is equipped with an Apogee AP7p CCD camera that uses 
a back-illuminated SITE 502A chip. Mrk 421 was observed during the 
period from 2003 November 8 to 2004 June 11. The data were taken with 
an uncalibrated Cousins R filter (designed by Bessell and manufactured 
by Omega). The photometric measurements are differential (with respect
to the comparison star 4 in Villata et al. 1998). Aperture photometry 
was performed with custom software. The aperture used is of 10\arcsec\ 
in diameter. Data points were obtained from averaging over between four 
and six 2 minute exposures. The typical statistical error on the relative 
photometry of each data point is 0.02 in magnitude. The seeing-induced 
errors or other systematic errors were not taken into account. To cross
check results from the two sets of measurements, we scaled up the Boltwood 
values (by adding a constant to the measured differential magnitudes) so 
that they agree, on average, with the FLWO fluxes for the overlapping 
time period. We found that the observed variation patterns agree quite
well between the two data sets. This also implies that any variability 
caused by systematic effects must be small compared with the intrinsic
variability of the source.

\subsection{Radio observations}

We observed Mrk 421 frequently with the 26-meter telescope at the 
University of Michigan Radio Astronomy Observatory (UMRAO) and the 
13.7-meter Mets\"{a}hovi radio telescope at the Helsinki University of
Technology. The UMRAO telescope is equipped with transistor-based 
radiometers operating at center frequencies of 4.8, 8.0, and 14.5 GHz; 
their bandwidths are 560, 760, and 1600 MHz, respectively. All three 
frequencies utilize rotating, dual-horn polarimeter feed systems, which 
permit both total flux and linear polarizations to be measured using an 
ON-OFF observing technique at 4.8 GHz and an ON-ON technique (switching 
the target source between the two feed horns closely spaced on the sky) at
the other two frequencies. The latter technique reduces the contribution 
of tropospheric interference by an order of magnitude. Observations of 
Mrk 421 were intermixed with observations of a grid of calibrator sources: 
3C 274 is the most frequently used calibrator, except during a period 
each fall when the sun is within 15$^\circ$ of the calibrator.  The flux 
scale is set by observations of Cas A (e.g. see Baars et al. 1977). 
Details of the calibration and analysis techniques are described in 
Aller et al. (1985).

The Mets\"{a}hovi observations were carried out as part of a long-time 
monitoring program. The observations were made with dual-horn receivers 
and the ON-ON technique, and they covered the 22 and 37 GHz bands. The 
flux measurements have been calibrated against DR 21 (with the adopted 
fluxes 19.0 
and 17.9 Jy at 22 and 37 GHz, respectively). The full description of the 
receiving system can be found in Ter\"{a}sranta et. al. (1998). The data 
were obtained during the period from 2003 January 1 to 2004 June 28.

\section{Results}

\subsection{Light Curves}

Figure~1 shows the light curves of Mrk 421 in the representative bands
covered by the campaign. The source was relatively quiet during the 
2002/2003 season, although it clearly varied at X-ray and TeV energies. 
The largest TeV flare occurred around MJD 52728 (= 30 March 2003), reaching 
a peak count rate of nearly 4 $\gamma\mbox{ }min^{-1}$ and lasting for 
about a week. The TeV flare was accompanied by a flare in X-rays, which 
reached a peak count rate of 
$\sim$65$\mbox{ }cts\mbox{ }s^{-1}\mbox{ }PCU^{-1}$ (or about 24 mCrab). 
There is no apparent time offset between the X-ray and gamma-ray flares.

Mrk 421 became much more active in the 2003/2004 season, with several major 
flares observed. In particular, an usually large flare (or outburst) took 
place near the end of the campaign, with the source reaching peak fluxes of 
$\sim$135 mCrab in X-rays, as seen by the ASM/RXTE,\footnote{based on data
from the MIT archive at http://xte.mit.edu/asmlc/srcs/mkn421.html\#data} 
and $\sim$3 Crabs in the TeV band, respectively. 
An expanded view of this flaring episode is shown in Figure~2. The flare 
lasted for more than two weeks (from $<$ MJD 53104 to roughly MJD 53120), 
although its exact duration is difficult to quantify due to the presence 
of a large data gap between MJD 53093 and MJD 53104. Interestingly, 
during this giant flare, the TeV emission appears to reach the peak much 
sooner than the X-ray emission. Although the X-ray light curve is not
as densely sampled as the gamma-ray light curve, the fact that the X-ray
measurements were made at the {\em same} time as the corresponding 
gamma-ray measurements (see Fig.~2) makes it quite unlikely that the 
difference in the rise time between the two bands is caused by some
sampling bias.

The light curves also show that Mrk 421 varies much less at optical and
radio wavelengths. The variability amplitude in the R band is less than 
about 0.9 magnitude; the variability is even less obvious in the radio 
bands due to relatively large measurement errors. There is no apparent 
correlation between the optical and radio bands or between either radio 
or optical band and the high-energy (X-ray and TeV) bands. For 
completeness, Figure~3 shows (FLWO) optical and radio light curves for 
all of the bands covered. The figure shows highly correlated variability 
among the optical bands, while the situation is not nearly as clear among 
the radio bands, due both to large measurement errors and sparse coverages.

\subsubsection{Energy Dependence of the Source Variability}

To quantify the energy dependence of variability amplitudes, we computed 
the so-called normalized variability amplitude (NVA) for each spectral band. 
The NVA is defined as (Edelson et al. 1996):
\[
NVA \equiv \frac{{({\sigma_{tot}}^2 -{\sigma_{err}}^2)}^{1/2}}{\bar{F}},
\]
where $\bar{F}$ represents the mean count rate or flux, $\sigma_{tot}$ 
the standard deviation, and ${\sigma_{err}}$ the mean measurement 
error in a given spectral band. 

To facilitate comparison of the NVAs for different bands, we only used
light curves that are relatively well sampled and cover the entire
2003/2004 season (see Fig.~1). Consequently, the results are obtained 
only for the following spectral bands: 14.5 GHz, R, X-ray, and gamma-ray.
We first rebinned the selected light curves with the same bin size. To 
examine variabilities on different timescales, two different bin sizes 
were used: 1 day and 7 days. We then computed the NVAs from the rebinned
light curves. The results are shown in Figure~4. The variability of
Mrk 421 shows a general increasing trend toward high energies. The source
is highly variable in the X-ray and TeV bands, with the NVA reaching up 
to about 65\% (with 1-day binning), but varies much less in the radio
and optical bands, with the NVA equal to $\sim$11\% and $\sim$16\%, 
respectively, almost independent of the binning schemes used. 

There are several caveats in the cross-band comparison. First of all, the 
density of data sampling is quite different for different bands. For 
instance, the sampling is more sparse in the 14.5 GHz band than in any of 
the other bands. However, the under-sampling of the radio data would
probably only lead to an {\em under-estimation} of the radio NVA. Secondly, 
the measurement errors for the optical data are probably underestimated, 
due to possible systematic effects caused by, e.g., the presence of bright 
stars in the field. This would result in an {\em over-estimation} of the 
optical NVA. Finally, the measured optical flux includes contribution from 
the host galaxy, which is at a $\sim$15\% level (Nilsson et al. 1999). 
Since it only affects the average flux, not the absolute variability 
amplitude, the optical NVA should be about 15\% higher, which is a small 
correction. Therefore, the conclusion that the NVA is comparable in the 
radio and optical bands seems secure. 

\subsubsection{X-ray--TeV Correlation}

From the light curves (Fig.~1), we can see that the X-ray and TeV 
variabilities of Mrk 421 are roughly correlated, although they are clearly 
not always in step. Figure~5 shows the X-ray and gamma-ray count rates for 
all simultaneous measurements. Although a positive correlation between the 
rates seems apparent, it is only a loose one. We should note that the 
dynamical range of the data is quite large ($\sim$30 in both energy bands) 
which is important for studying correlative variability of the source. 

To be more rigorous, we computed the Z-transformed discrete correlation 
function (ZDCF; Alexander 1997) from light curves in the two bands. The 
ZDCF makes use of the Fisher's z-transform of the correlation coefficient 
(see Alexander 1997 for a detailed description). Its main advantage over 
the more commonly used DCF (Edelson \& Krolik 1988) is that it is more 
efficient in detecting any correlation present. Figure~6 shows the ZDCF (in 
1-day bins) derived from the 2002/2003 data set. The ZDCF seems to peak at 
a negative lag. Fitting the peak (in the narrow range of -7--7 days) with a 
Gaussian function, we found its centroid at $-1.8 \pm 0.4$ days, which is
of marginal significance. If real, the result would imply that the X-ray 
variability {\em leads} the gamma-ray variability. Other ZDCF peaks are 
most likely caused by the {\em Whipple} observing pattern, such as the 
quasi-periodic occurrences of the dark moon periods. Similarly, we computed 
a ZDCF for the 2003/2004 data set. The results are shown in Figure~7. In 
this case, the main feature is very broad and significantly skewed toward 
{\em positive} lags. The feature appears to be a composite of multiple 
peaks, although large error bars preclude a definitive conclusion. A 
positive ZDCF peak means that the X-ray emission {\em lags} behind the 
gamma-ray emission. We checked the results with different binning schemes 
and found no significant changes. 

The detection of X-ray {\em lags} in the 2003/2004 data set should not 
come as a total surprise, because we have already seen (from Fig.~2) that 
the X-ray flux rose more slowly than the gamma-ray flux during the 2004
giant flare. The difference in the rise times can be the cause of the 
broad ZDCF peak that is skewed toward positive lags. To check that, we 
excluded the flare from the X-ray and gamma-ray light curves (by removing 
all data points after MJD 53100; see Fig.~1) and computed a new ZDCF. The 
results are also shown in Fig.~7 (bottom panel) for 
a direct comparison. There is a narrow peak at around -1 day. Fitting it 
(in the range of -7--3 days) with a Gaussian function yields the centroid 
at $-1.2 \pm 0.5$ days, which is not inconsistent the measured value for 
the 2002/2003 data set though it is even less significant statistically. 
The features at around +7 and -15 days (which are also discernable in the 
overall ZDCF) are, again, most likely caused by the {\em Whipple} observing 
pattern for the season. Despite the complications, it is almost certain, by 
comparing the two ZDCFs, that the X-ray lags in the 2003/2004 data set are 
indeed associated with the difference between the X-ray and gamma-ray rise 
times of the giant flare.

To investigate the effects of data gaps (which are present both in the 
X-ray and gamma-ray data) on ZDCF, we did the following experiment. We 
created three light curves: 
the actual X-ray light curve of Mrk 421 in the 2003/2004 season({\em lc1}), 
an artificial light curve ({\em lc2}) made by shifting {\em lc1} by +8 days, 
and another artificial light curve ({\em lc3}) made by modulating {\em lc2} 
with the {\em Whipple} sampling pattern. Figure~8 shows the ZDCFs between 
{\it lc1} and {\it lc2} and between {\it lc1} and {\it lc3} separately. 
The artificially-introduced 8-day lag is easily recovered in both cases. 
We also shifted {\em lc1} by different amounts and the lags are always
recovered correctly. Therefore, the coverage gaps associated with the 
{\em RXTE} and {\em Whipple} monitoring campaigns do not wash out intrinsic 
offsets between the X-ray and gamma-ray light curves, although the effects 
of the data gaps on the shape of the ZDCFs are measurable. 

\subsubsection{X-ray and TeV Flares}

Examining the X-ray and TeV light curves more closely, we noticed that
some of the TeV flares have no {\em simultaneous} X-ray counterparts
(or counterparts at long wavelengths). They are, therefore, similar 
to the reported ``orphan'' TeV flare in 1ES1959+650 (Krawczynski et al.
2004). Figure~9 shows an example of the phenomenon. The TeV flare peaks 
at almost 8 $\gamma\mbox{ }min^{-1}$ at around MJD 53033.4 when the 
X-ray flux is low. It is interesting to note, 
however, that the source is clearly variable in X-rays during this time 
period and that the X-ray flux seems to have peaked about 1.5 days before 
the TeV flux. Therefore, the TeV flare might not be a true orphan event; 
it might simply lag behind its X-ray counterpart. Alternatively, it is 
also possible that the TeV flare is a composite of two sub-flares, 
with one being the counterpart of the X-ray flare and the other a true, 
lagging orphan flare. The sparse sampling of the data prevents us from 
drawing a definitive conclusion in this regard. We note the remarkable
similarities between the TeV flare shown in Fig.~9 and the reported
``orphan'' flare in 1ES1959+650 (see Fig.~4 in Krawczynski et al. 2004), 
including similar variation patterns in X-rays. 

From the X-ray and TeV light curves, we also detected flares on much
shorter timescales. Figure~10 shows examples of sub-hour X-ray flares.
The most rapid X-ray flare detected lasted only for $\sim$20 minutes and 
shows substantial sub-structures, implying variability on even shorter
timescales. Only on one occasion was a sub-hour X-ray flare detected 
during a (longer-duration) gamma-ray flare. No counterpart is apparent 
at TeV energies (see Fig.~10), although the error bars on gamma-ray 
measurements are quite large. We note that, if a strong rapid TeV flare, 
for example, like the one detected by Gaidos et al. (1995), had occurred, 
we should have easily detected it, given the improved instrumentation. 
The rapid X-ray flare shown in Fig.~10 is relatively weak, with a 
peak-to-peak amplitude of only about 5\% of the average flux level. The 
data do not allow any direct comparison on long timescales, due to the 
short exposure of the X-ray observation. 

\subsection{Spectral Energy Distributions}

We first divided the {\em RXTE} observations into 8 groups based on the 
X-ray count rate of Mrk 421, with an increment of 
20 $cts\mbox{ }s^{-1}\mbox{ }PCU^{-1}$. For this work,
we focused on three of the groups: 
low (0--20 $cts\mbox{ }s^{-1}\mbox{ }PCU^{-1}$), 
medium (40--60 $cts\mbox{ }s^{-1}\mbox{ }PCU^{-1}$), 
and high (140--160 $cts\mbox{ }s^{-1}\mbox{ }PCU^{-1}$), with the 
average count rate 
increasing roughly by a factor of three from low to medium and from medium
to high. Then, for each observation in a group we searched for an 
observation at TeV energies that was made within an hour. Only matched
pairs were kept for further analysis. We ended up with a total of 16, 
9, and 3 pairs in the low-flux, medium-flux, and high-flux group, 
respectively. It turns out that the low-flux
group consists of observations taken between 2003 March 8 and May 3,
the medium-flux group between 2004 January 27 and March 26, and the
high-flux group between 2004 April 16--20.

For each group, we proceeded to construct a flux-averaged SED at X-ray
and TeV energies. The flux-averaged X-ray spectrum can 
be fitted satisfactorily by a power law with an exponential roll-over 
(with reduced $\chi^2 \lesssim 1$). We should note that we added a 
1\% systematic uncertainty to the X-ray data for spectral analysis
(which is a common practice), to take into account any residual 
calibration uncertainty. Also, we fixed the hydrogen column density at 
$1.38\times 10^{20}\mbox{ }cm^{-2}$ (Dickey \& Lockman, 1990). The 
best-fit photon index is about $2.51^{+0.03}_{-0.05}$, 
$2.38^{+0.02}_{-0.02}$, and $1.99^{+0.02}_{-0.02}$ for the low-, medium-, 
and high-flux group, respectively, and the roll-over energy about 
$26^{+3}_{-5}$, $32^{+3}_{-3}$, and $32^{+2}_{-2}$ keV. Therefore, the 
X-ray spectrum of Mrk 421 hardens toward high fluxes. Using the best-fit 
model we then derived the X-ray SED for each data group.

For the gamma-ray spectral fits, a bin size of 0.1667 in $\log_{10}(E)$ 
was adopted for the medium- and high-flux groups and a wider bin size 
of 0.4 for the low-flux group. As shown in Fig.~12, the low-flux SED 
still has very large error bars. The first energy bin corresponds to an 
energy of $\sim$260\,GeV. The gamma-ray spectra can all be satisfactorily 
fit by a power law, with a photon index of $2.84\pm0.58$, 
$2.71\pm0.15$, and $2.60\pm0.11$ for the low-, medium-, and 
high-flux groups, respectively. The errors bars only include statistical
contributions. For the purpose of comparison with some of the published 
results, we also fitted the spectra with a cut-off power law (see, e.g.,
Krennrich et al. 2002) but with the cut-off energy fixed at 4.96 TeV.
The photon index is $2.73\pm0.56$, $2.40\pm0.18$, and $2.11\pm0.14$
for the low-, medium-, and high-flux groups, respectively. Like the X-ray 
spectrum, therefore, the TeV spectrum also seems to harden toward high 
fluxes, although the uncertainty here is quite large.

Finally, we searched for radio and optical observations that fall in 
one of the groups and computed the average fluxes to complete the
SED for the group. Given the fact that Mrk 421 did not vary as 
significantly at these wavelengths, we believe that the derived SEDs 
are quite reliable. Figure~12 summarizes the results.

\subsection{Spectral Modeling}

We experimented with using a one-zone SSC model (see Krawczynski et al. 
2004 for a detailed description of the code, which we revised to use 
the adopted cosmological parameters) to fit the measured flux-averaged 
SEDs. Briefly, the model calculates the SED of 
a spherical blob of radius $R$. The blob moves down the conical 
jet with a Lorentz factor $\Gamma$. The emitting region is filled 
with relativistic electrons with a broken power-law spectral 
distribution: $S_e \propto E^{-p_1}$, for $E_{min}<E<E_{b}$, and $S_e
\propto E^{-p_2}$, for $E_{b} \le E<E_{max}$ (although the code does
not evolve the electron spectrum self-consistently). The model accounts 
for the attenuation of the very-high-energy $\gamma$-rays by the 
diffuse infrared background (as modelled by MacMinn \& Primack 1996) . 

We first found an initial ``best fit'' to each SED by visual inspections. 
We then performed a systematic grid search around the ``best fit'' that 
involves the following parameters: magnetic field $B$ in the range of 
0.045--0.45 G, 
Doppler factor $\delta$ in 10--20, $p_1$ in 1.8--2.2, $p_2$ in 2.9--3.7,
$log{E_b}$ in 9.9--12.2, $log{E_{max}}$ in 9.9--12.2, and the 
normalization ($w_e$) of $S_e$ in 0.00675--0.44325 $ergs\mbox{ }cm^{-3}$. 
Variability constraints (see the next section) were taken into account 
in the choice of some of the parameter ranges. All other parameters in 
the model (e.g., $R$) were fixed at the nominal values determined by the 
visual inspections. We found roughly where the global $\chi^2$ minimum 
lies through a coarse-grid search, and then conducted a finer-grid search 
through much smaller parameter ranges around the minimum to find the best 
fit. Figure~11 shows the results for the high-flux group. It is apparent
that the model severely underestimates the radio and optical fluxes. Large 
deviations are also apparent at TeV energies. We should stress that our
grid searches are, by no means, exhaustive. However, the results should be 
adequate for revealing gross discrepancies between the model and the data. 

To investigate whether the fit could be improved by introducing 
additional zones (which are assumed to be independent of each other), 
we introduced a new zone to account for the radio fluxes and another for 
the optical fluxes. To be consistent with the observed variability, the 
additional zones must be placed much further down the jets. Figure~12 
shows fits to the SEDs with such a three-zone SSC model. This ad-hoc 
approach does seem to yield a reasonable fit to the data. In such a 
scenario, the observed variability at X-ray and TeV energies would be 
associated with zones close to the central black hole, while radio
and optical emission are expected to vary on longer timescales further
down the jet.

\section{Discussion}

\subsection{X-ray--TeV Correlation}

The success of the standard SSC model lies partly in the fact that it 
provides a natural explanation for the correlation between the X-ray and 
gamma-ray emission. However, we found that the correlation is not as
tight as one might naively expect from the SSC model. It seems unlikely 
that the loose correlation is caused by the choice of 
spectral bands used in the cross-correlation analysis, given how broad 
the {\em PCA} and {\em Whipple} passing bands are. In fact, both the 
derived variability amplitudes (Fig.~4) and SEDs (Fig.~12) indicate that 
the X-ray and gamma-ray photons are likely to have originated from the 
same population of electrons in the context of the SSC model.

In the standard SSC model, X-ray photons are Compton upscattered to produce 
gamma-ray photons, so an X-ray lead is naturally expected. However, the 
characteristic timescale of the SSC process would be much too short to 
account for the X-ray lead of 1--2 days (see Fig.~6), although we cannot 
be sure about whether the lead is actually real, due to large uncertainties
and possible systematic effects. What is certain is that the SSC model 
cannot explain the difference between the X-ray and gamma-ray rise times 
of the 2004 giant flare. It is conceivable that the flaring episode could 
have started with an orphan TeV flare and followed by a pair of 
simultaneous X-ray and TeV flares. If this is the case, it would be 
opposite to the theoretical scenario recently put forth by B\"{o}ttcher 
(2004). In the hadronic models, both X-ray lag and lead could, in principle,
occur, even at the same time, depending on the relative roles of the primary 
and secondary electrons. However, it would also seem challenging for these
models to account for the different X-ray and gamma-ray rise times of the
2004 giant flare or to quantitatively explain the long X-ray lead (of 1--2
days) if it is real.

\subsection{Variability Constraints}

Combined with the measured SED, rapid X-ray flares pose severe constraints 
on some of the physical properties of the flaring region, in a relatively 
model-independent manner, because the X-ray emission from Mrk~421 is 
almost certainly of synchrotron origin. The most rapid X-ray flare detected 
during our campaign has a duration of about 20 minutes and reaches a peak 
amplitude of $\sim$15\% of the (local) non-flaring flux level (see Fig.~10). 
Therefore, the size of the flaring region must satisfy:
$l' \lesssim c t_{flare} \delta/(1+z) = 3.6\times 10^{14} \delta_1\mbox{ }cm$,
where $t_{flare}$ is the duration of the flare ($=1200$ s), $\delta$ is the
Doppler factor of the jet ($\delta = 10\delta_1$), and $z$ is the redshift 
of Mrk 421 ($z = 0.031$). Here and in the remainder of the paper, we use
the primed symbols to denote quantities in the co-moving frame of the jet
and unprimed ones corresponding quantities in the frame of the observer.
It is interesting to note that the derived upper limit is already comparable 
to the radius of the last stable orbit around the $2\times 10^8 M_{\odot}$ 
black hole in Mrk 421 (Barth et al. 2003), for $\delta \sim 10$. Since 
the peak flux of the flare is a significant fraction of the steady-state 
flux, the size of the flaring region is probably comparable to the lateral 
extent of the jet (to within an order of magnitude). If the jet originates 
from accretion flows, as is often thought to be the case, the result would 
also represent an upper limit on the inner boundary of the flows.

The decaying time of the most rapid X-ray flare (about 600 s) sets a firm 
upper limit on the synchrotron cooling time of the emitting electrons. 
The cooling time is given by (Rybicki \& Lightman 1979), 
$\tau'_{syn} \approx 6 \pi m_e c/\sigma_T \gamma'_b B'^2$, where
$m_e$ is the electron rest mass, $\sigma_T$ is the Thomson cross
section, $B'$ is the strength of the magnetic field in the region, and 
$\gamma'_b$ is the characteristic Lorentz factor of those
electrons that contribute to the bulk of the observed X-ray emission. 
By requiring $\tau'_{syn} < t_d \delta$, where $t_d$ is the measured
decaying time of the flare ($\approx 600 s$), we derived a lower 
limit on $B'$, $B' > 1.1 \delta_1^{-1/2} {\gamma'}_{b,5}^{-1/2} \mbox{ }G$, 
where $\gamma'_b = 10^5\gamma'_{b,5}$. From modeling the SEDs, we found 
$2 \times 10^5 \le \gamma'_b \le 4 \times 10^5$. It should be noted that
the limits derived are only appropriate for the region that produced
the 20-min X-ray flare and that not all flares are necessarily 
associated with the same region. In our attempt to model the SEDs 
(\S~3.2), we only require that the model parameters be consistent 
with variability timescales of hours. Results from more sophisticated 
modeling will be presented elsewhere.

Further constraints can be derived from the detected TeV flares. The 
fact that we detect TeV photons from Mrk 421 requires that the opacity
due to $\gamma\gamma \rightarrow e^{+} e^{-}$ must be sufficiently
small near the TeV emitting region(s). This requirement leads to
(Dondi \& Ghisellini 1995):
\[\delta \geq \left[ \frac{\sigma_T d^2_L}{5hc^2} \frac{F(\nu_t)}{t_{var}} \right]^{1/(4+2\alpha)}, \] 
where $d_L$ is the luminosity distance, $t_{var}$ is the TeV flux doubling 
time, $\alpha$ and $F(\nu_t)$ are the local spectral index and the 
energy flux, respectively, at the target photon frequency, 
at which the pair production cross section peaks,
\[ {\nu} {\nu_{t}} \sim \left( \frac{m_e c^2}{h} \frac{\delta}{1+z} \right)^2, \]
where $\nu$ is the frequency of the gamma-ray photon. For $h\nu \sim 1$ TeV,
we have $\nu_t \sim 6 \times 10^{15}$ Hz. From the high-flux SED (Fig.~12), 
we found $\alpha \sim 0.5$ and 
$\nu_t F_{\nu_t}\sim 1\times 10^{-10}\mbox{ }erg\mbox{ }cm^2\mbox{ }s^{-1}$.
Given $d_L = 129.8$ Mpc and $t_{var} \simeq 1$ hour (see Fig.~10), we 
derived a lower limit on the Doppler factor, $\delta \gtrsim 10$.

\subsection{``Orphan'' TeV Flares}

Since TeV emission is the consequence of inverse-Compton scattering of 
(synchrotron) X-ray photons by the electrons themselves in the standard SSC 
model, a flare at TeV energies must be {\em preceded} by an accompanying 
flare at X-ray energies. While the model might still be able to accommodate 
the presence of orphan X-ray flares, the presence of orphan TeV flares seems 
problematic. On the other hand, the ``orphan'' TeV flares in Mrk 421 (as 
shown Fig.~9) and in 1ES1959+650 (Krawczynski et al. 2004) might not be 
orphan events after all. In both cases, the gamma-ray flares seem to be 
preceded by X-ray flares, which could be attributed to the SSC process 
although the X-ray lead (of 1--2 days) would seem too long. Alternatively, 
it might be possible to attribute the X-ray lead to some sort of feedback 
between different emission regions in the jet (i.e., an orphan X-ray flare 
in one region triggers an orphan TeV in another). 

Genuine orphan TeV flaring would not be easy to understand in the hadronic 
models either, despite looser coupling between the X-ray and TeV emission, 
unless the injected e/p ratio changes significantly from flare to flare. 
Any change in the proton population will likely be accompanied by a change 
in the electron (primary or secondary) population, so TeV and X-ray flares
should both occur as a result. Interestingly, orphan TeV flares may be 
understood in a hybrid scenario in which protons are present in the jet in 
substantial quantities but are not necessarily dominant compared to the 
lepton component (B\"{o}ttcher 2004). In this case, an orphan TeV flare is 
associated with (and {\em }follows) a pair of simultaneous X-ray and TeV 
flares that originate in the standard SSC process. The synchrotron (X-ray) 
photons are then reflected off some external cloud and return to the jet 
(following a substantial delay with respect to the initial flares). The 
returned photons interact with the protons in the jet to produce pions and 
subsequent $\pi^0$ decay produces TeV photons in the flare. The model was 
shown to be able to account for the ``orphan'' TeV flare observed in 
1ES1959+650 (B\"{o}ttcher 2004). It might also explain some of the similar 
flares in Mrk 421, for example, the one shown in Fig.~9, if it is a 
composite of two gamma-ray flares (see discussion in \S~3.1.3).

\subsection{Spectral Energy Distribution}

The SED of Mrk 421 varies greatly both at X-ray and TeV energies but 
only weakly at radio and optical wavelengths. The much reduced variability 
at long wavelengths is expected for short timescales, because of long 
synchrotron cooling times of the radio or optical emitting electrons. 
In other words, there is no reason to expect a tight correlation between 
the low-energy (radio or optical) emission and the high-energy (X-ray or 
TeV) emission. Of course, the lack of such a correlation could also be due 
to the fact that the low-energy emission originates in regions further 
down the jet. Between the radio and optical bands, we found that the NVAs 
of the source are comparable (see Fig.~4), 
which is somewhat surprising because the synchrotron cooling time of the 
electrons responsible for the optical emission is expected to be much 
shorter than that of those for the radio emission. On the other hand, this 
could be evidence for the presence of different populations of electrons 
that produce the emission in the radio and optical bands. The populations 
may be located in physically separated regions and may have different 
spectral energy distributions (e.g., $E_{max}$).

We found that both SED peaks move to higher energies as the luminosity 
of the source increases. Moreover, the X-ray spectrum becomes flatter, 
which seems to be common among high-frequency-peaked BL Lacs (e.g., 
Giommi et al. 1990; Pian et al. 1998; Xue \& Cui 2004). Similar spectral 
hardening has also 
been see at TeV energies (Krennrich et al. 2002; Aharonian et al. 2002). 
In our data set, we have seen some indication of spectral hardening but 
we cannot be certain because of the low statistics of the data. The 
evolution of the SED could be driven by a hardening in the electron energy 
distribution and/or an increase in the strength of the magnetic field at 
high luminosities.

\section{Summary}

In this paper, we have presented a substantial amount of multi-wavelength 
data that have been collected on Mrk 421 as the result of a long-term 
monitoring campaign. The large dynamical range of the data has allowed 
us to carry out some detailed investigations on variability timescales, 
cross-band correlation, and spectral variability. The main results from 
this work are summarize as follows:
\begin{itemize}
\item We have shown that the emission from Mrk 421 varied greatly at both 
X-ray and gamma-ray energies and that the variabilities in the two 
energy bands are generally correlated. This is consistent with some of 
the earlier results but the dynamical range of the data and statistics 
are much improved here. Equally important is the finding that the 
correlation is only a fairly loose one (see Figures 5--7). The emission 
clearly does not always vary in step between the two bands.
\item We have discovered that the X-ray emission reached the peak days 
{\em after} the TeV emission during the giant flare in April 2004 (see 
Fig.~1). Such a difference in the rise times for the two energy bands 
poses a serious challenge to the standard SSC model, as well as the 
hadronic models. Whether it can be accommodated by some hybrid model 
remains to be seen. In addition, there is tentative evidence that the 
X-ray variability leads the gamma-ray variability by 1--2 days. If proven 
real, such a long X-rad lead would also be difficult for either the SSC 
or hadronic models to explain quantitatively. 
\item We have detected X-ray and TeV flares on relatively short timescales.
For example, the most rapid X-ray flare detected has the duration of 
only $\sim$20 minutes and a peak amplitude of $\sim$15\% of the local
non-flaring flux level. Although no similarly rapid TeV flares were
detected during our campaign, we did observe a TeV flare with duration 
of hours. Physical constraints on the flaring regions have been 
derived from the rapid X-ray and TeV variabilities.
\item We have seen TeV flares in Mrk 421 that are similar to the reported 
``orphan'' flare in 1ES1959+650. 
Genuinely orphan or not, this presents another serious challenge to the 
proposed emission models (see \S~4.3). If they are true orphan events, they 
may be explained by a hybrid model in which both electrons and protons 
are present in the jet in substantial quantities. On the other hand, if 
the preceding X-ray flare turns out to be the counterpart, the 1--2 day
X-ray lead would challenge the proposed emission models (see \S~3.1.3).
\item We have derived high-quality SEDs of Mrk 421 from simultaneous or 
nearly simultaneous observations at TeV, X-ray, optical, and radio energies.
The X-ray spectrum clearly hardens toward high fluxes; the gamma-ray 
spectrum also appears to evolve similarly (although we cannot be sure due
to large uncertainties of the data). A one-zone SSC model fails badly to 
fit the measured fluxes at the radio and optical wavelengths, and, to a 
lesser extent, also underestimates fluxes at the highest TeV energies. The 
introduction of additional zones improves the fit significantly.
\item Further evidence for the presence of multiple populations of emitting
electrons is provided by the comparable radio and optical NVAs (see Fig.~4
and discussion in \S~4.4).
\end{itemize} 

\acknowledgements
We acknowledge the technical assistance of E. Roache and J. Melnick. The 
VERITAS collaboration is supported by the U.S. Department of Energy, 
National Science Foundation (NSF), the Smithsonian Institution, NSERC 
(Canada), PPARC (UK), and Science Foundation Ireland. W. Cui and 
M. B\l{}a\.{z}ejowski also gratefully acknowledge financial support 
from NASA and thank Peter Biermann for useful discussions on the subject 
and comments on the manuscript. UMRAO is supported in part by funds from 
NSF and from the Department of Astronomy at University of Michigan.

\clearpage

\begin{figure}
\psfig{figure=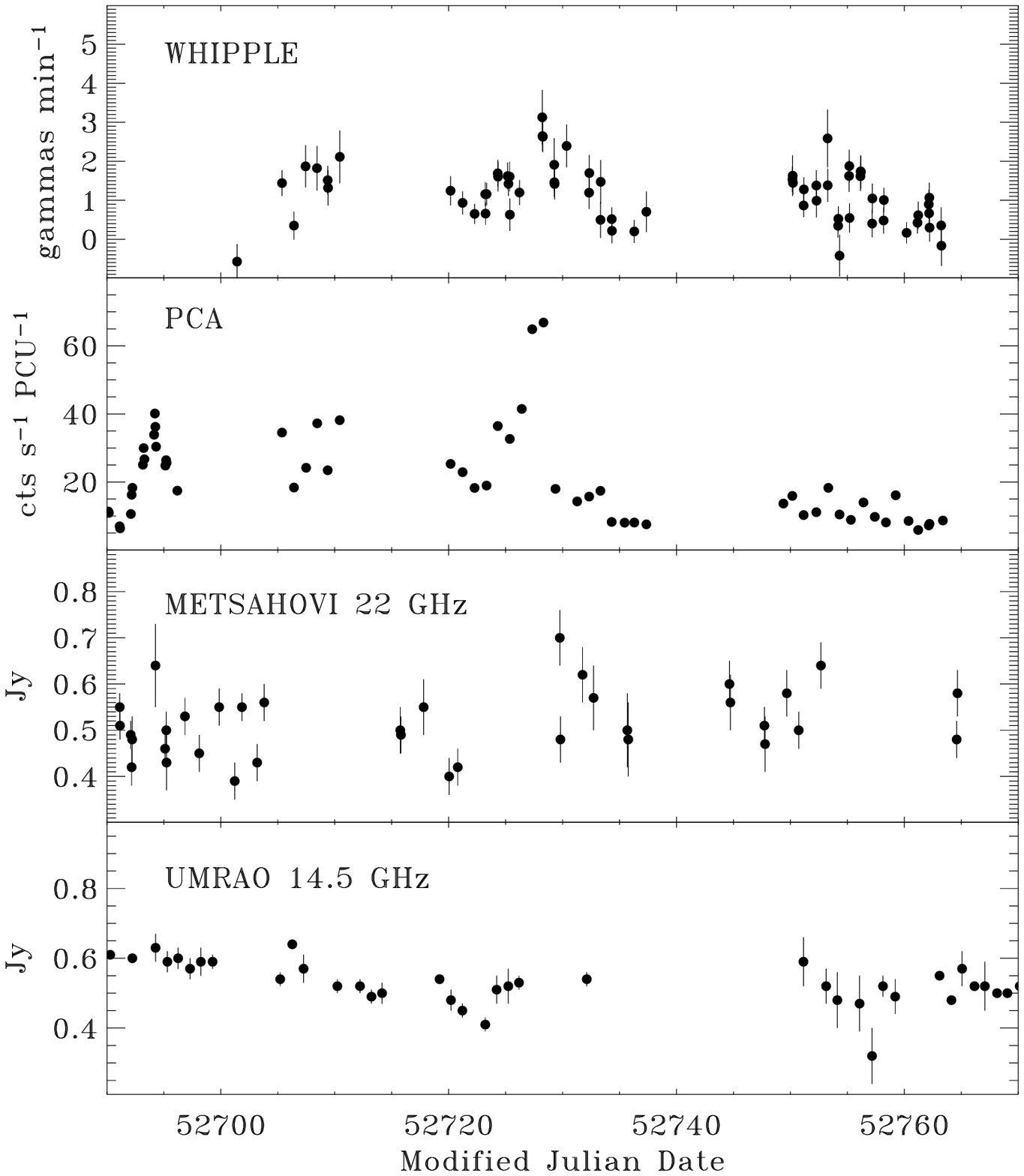,width=3.2in}
\psfig{figure=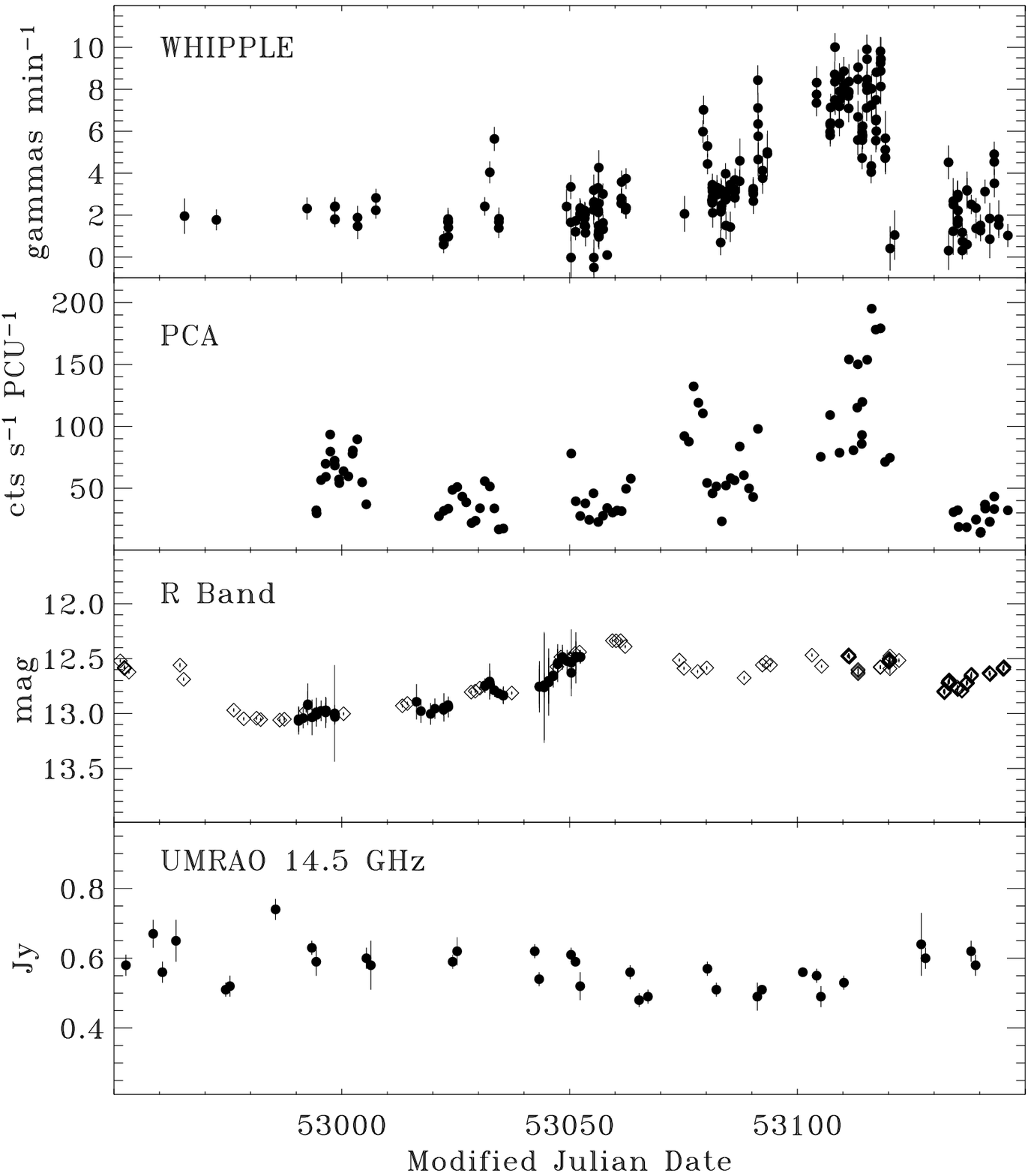,width=3.2in}
\caption{Multiwavelength light curves of Mrk 421. For clarity, the light 
curves are shown separately for the 2002/2003 ({\em left}) and 2003/2004 
({\em right}) {\em Whipple} observing seasons. For the R-band data, the FLWO 
points are shown in bullets and the Boltwood points in diamonds. Note 
that the gamma-ray rates have been corrected for zenith-angle changes, 
as well as for relative variations in the throughput of the telescope 
within each season. Also, the differential Boltwood measurements have been
scaled up to match the FLWO fluxes (see text). No optical data are 
available for the 2002/2003 season; the 22 GHz data are shown instead. }
\end{figure}

\begin{figure}
\psfig{figure=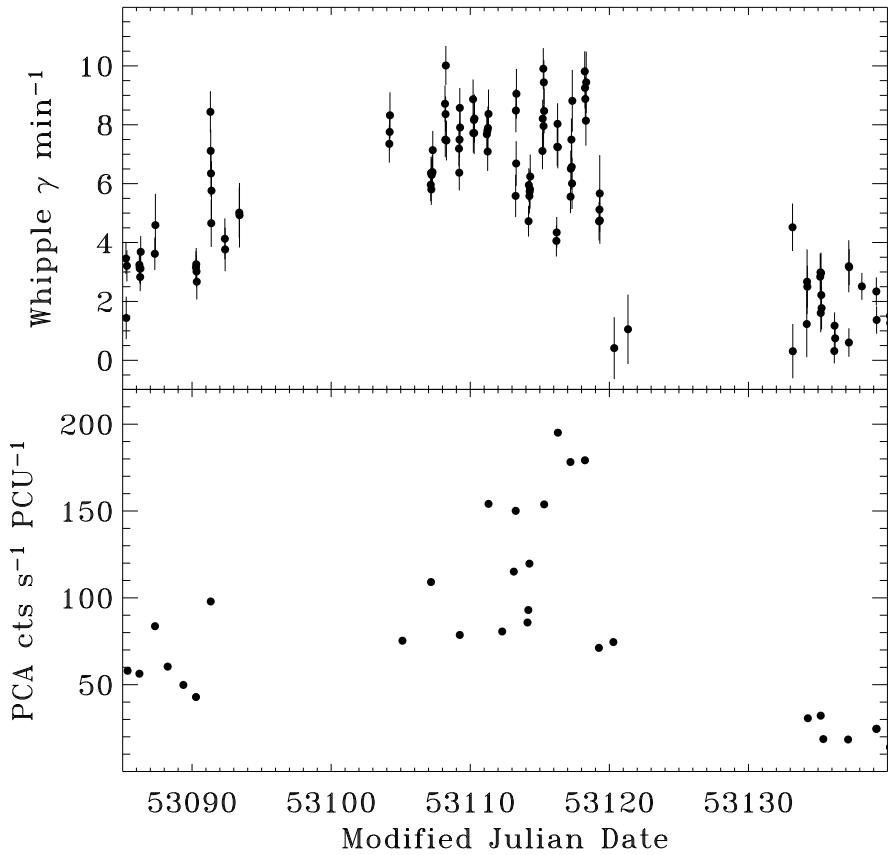,width=5in}
\caption{An expanded view of the major flaring episode in 2004 at both
gamma-ray ({\em Whipple}) and X-ray ({\em PCA}) energies. Note that on 
average the 
emission reaches the peak in gamma rays days {\em before} in X-rays. }
\end{figure}

\begin{figure}
\psfig{figure=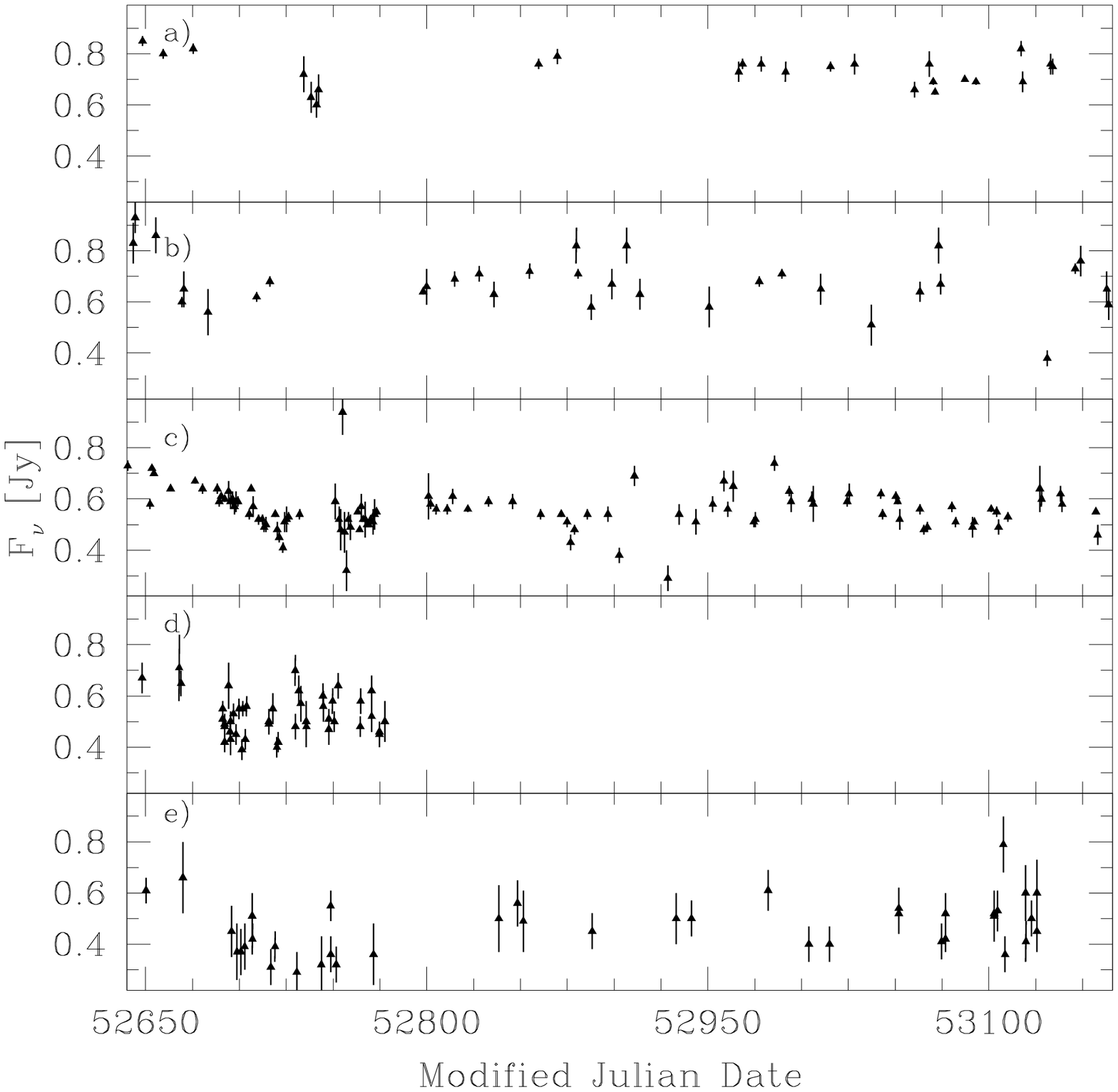,width=3.2in}
\psfig{figure=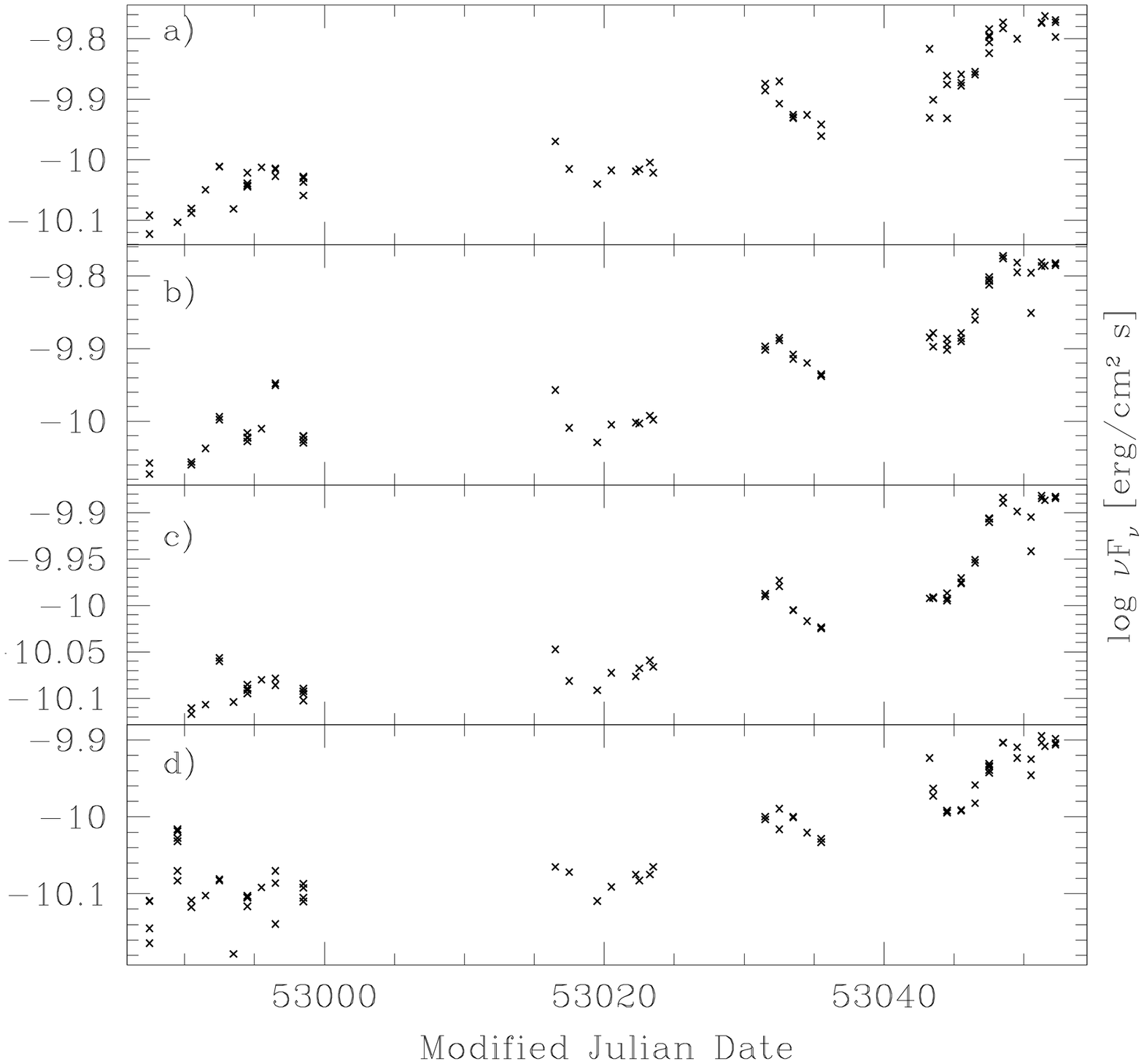,width=3.2in}
\caption{{\em (left)} Radio light curves of Mrk 421: a) 4.8GHz, b) 8.0GHz, 
c) 14.5GHz, d) 22GHz, and e) 37GHz. {\em (right)} Optical light curves of
Mrk 421: a) B band, b) V band, c) R band, and d) I band. Note that the
Boltwood R-band measurements are not shown here. }
\end{figure}

\begin{figure}
\psfig{figure=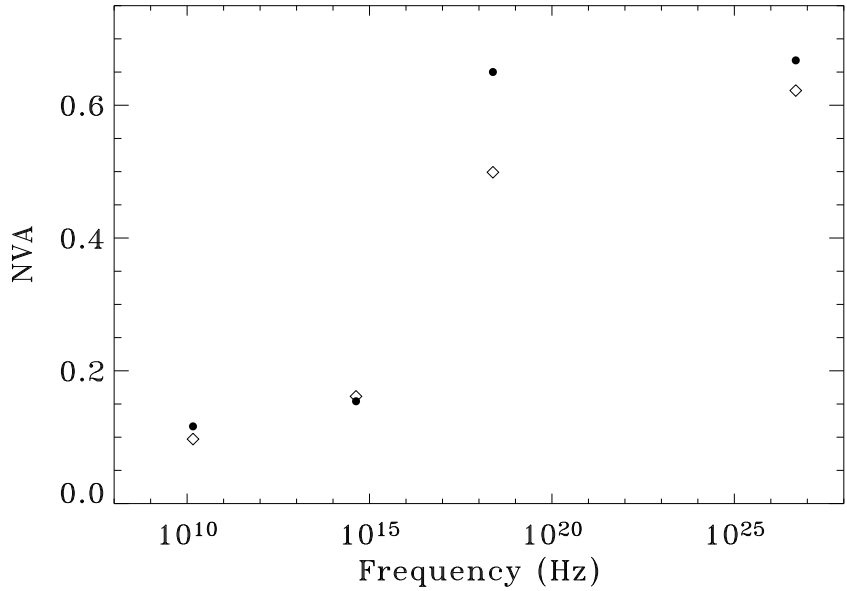,width=5in}
\caption{Normalized variability amplitudes in various spectral bands. 
The results from light curves with 1-day binning are shown in bullets 
and those with 7-day binning in diamonds. }
\end{figure}

\begin{figure}
\psfig{figure=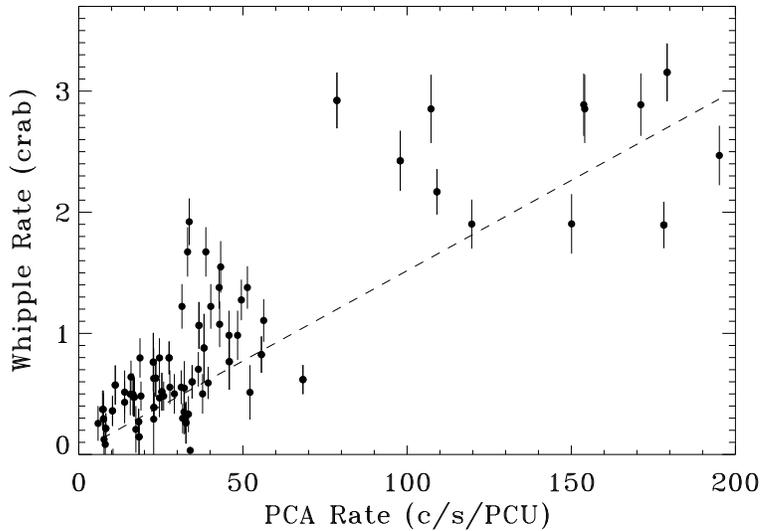,width=4.5in}
\caption{X-ray and gamma-ray count rates of Mrk 421. The figure shows
all pairs of X-ray and gamma-ray data points matched to within 15 
minutes. The dashed line shows the best linear fit, to guide the eye. 
Note the large deviations of many points from the fit. }
\end{figure}

\begin{figure}
\psfig{figure=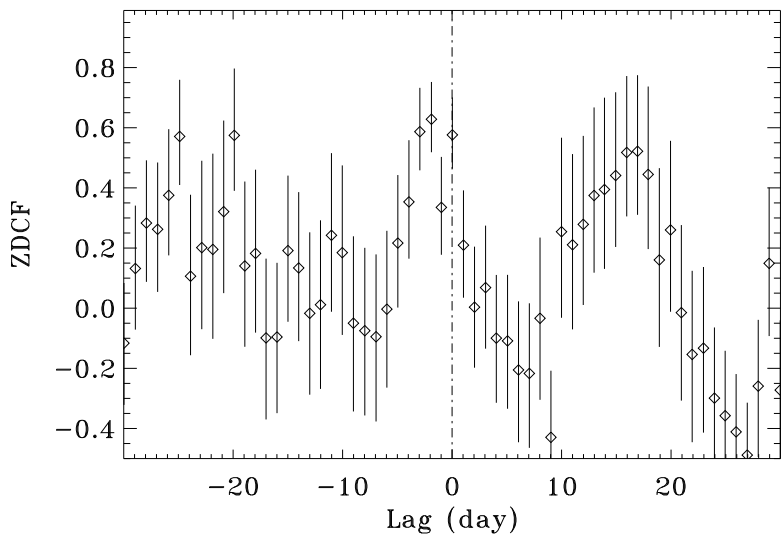,width=4.5in}
\caption{Z-transformed discrete correlation function between the X-ray 
and TeV light curves for the 2002/2003 season. }
\end{figure}

\begin{figure}
\psfig{figure=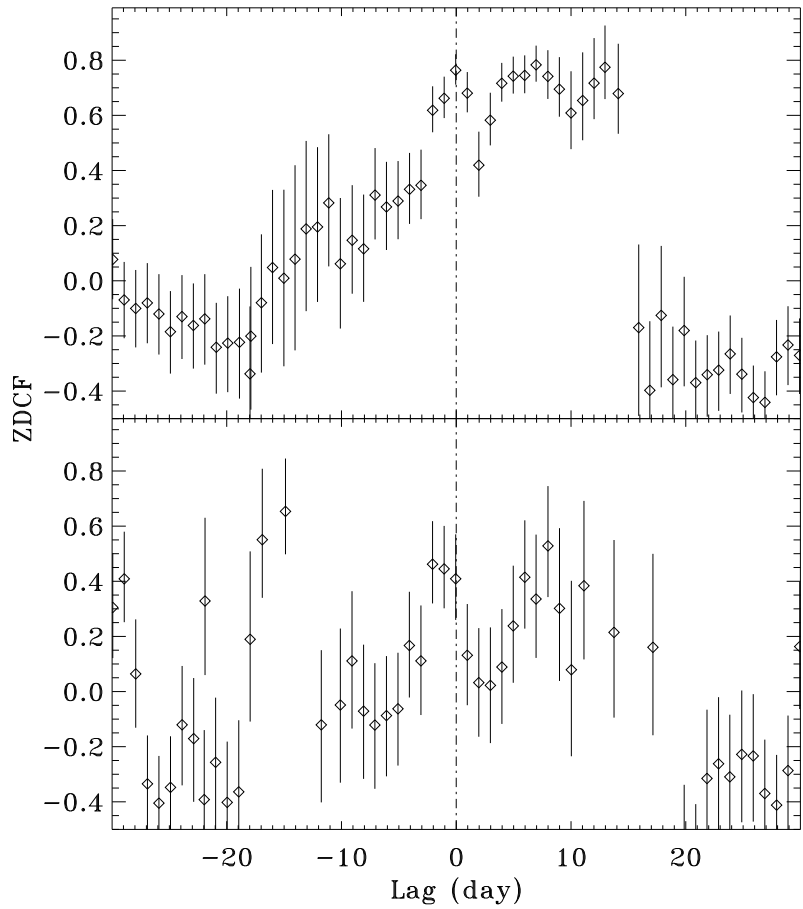,width=4.5in}
\caption{As in Fig. 6, but for the 2003/2004 season. Two ZDCFs are 
shown: {\em (top)} for the entire season and {\em (bottom)} for the
time period before the giant outburst (or MJD 53100; see Fig.~1). }
\end{figure}

\begin{figure}
\psfig{figure=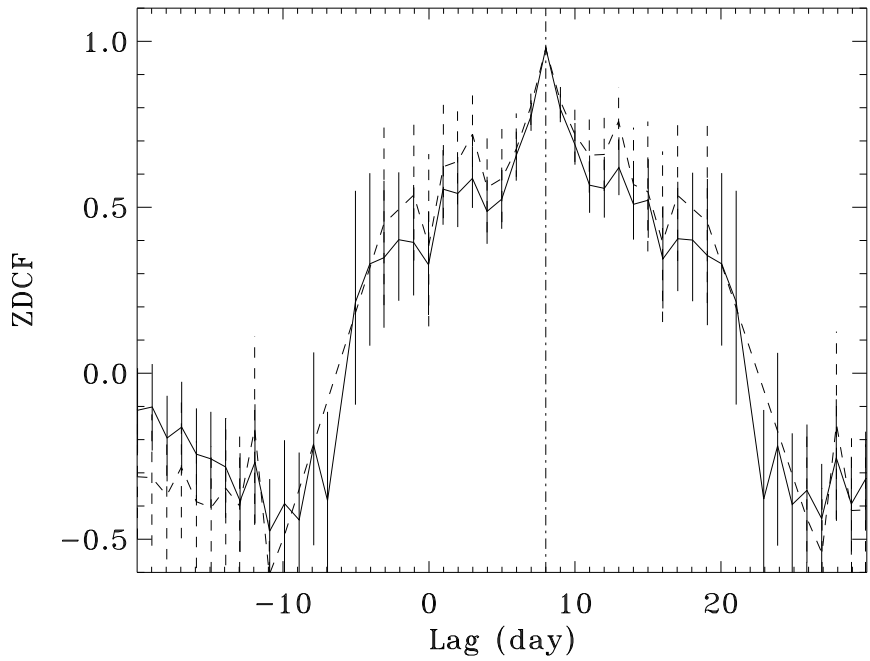,width=5in}
\caption{Simulated Z-transformed discrete correlation function between
{\em lc1} and {\em lc2} (in dashed line) and {\em lc1} and {\em lc3} 
(in solid line). See text for the definition of the light curves used.
({\em lc's}). Note that the ZDCF peaks exactly at the artificial 
+8 day lag in both cases.}
\end{figure}

\begin{figure}
\psfig{figure=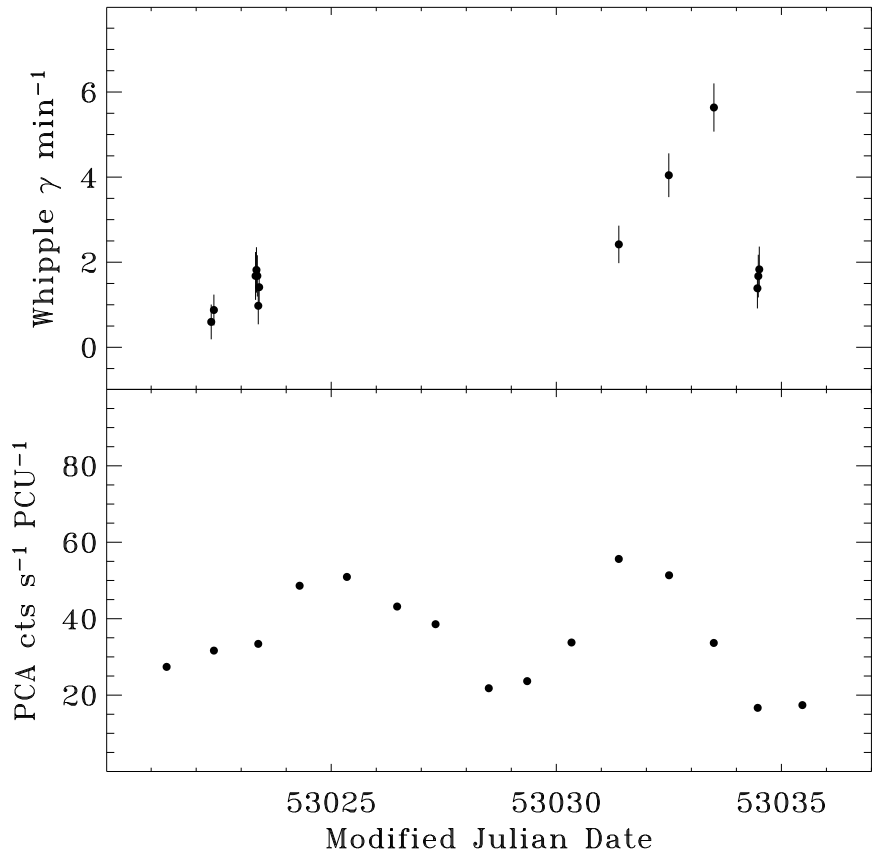,width=5in}
\caption{A TeV flare with no simultaneous X-ray counterpart: {\em (upper)} 
{\em Whipple} gamma-ray light curve and {\em (lower)} {\em PCA} X-ray 
light curve. 
Note the presence of significant X-ray variations during the period. }
\end{figure}

\begin{figure}
\psfig{figure=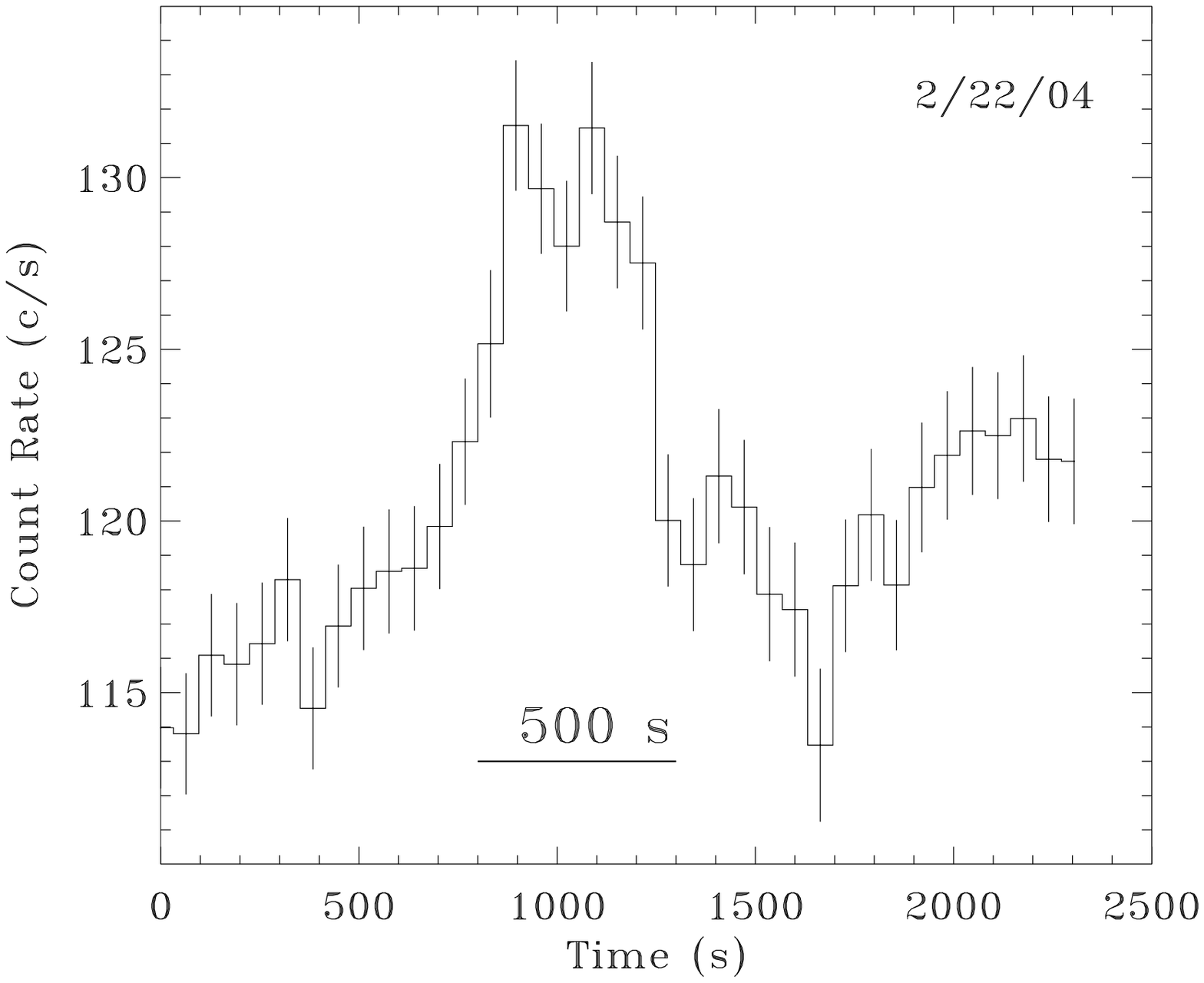,width=3.0in}
\psfig{figure=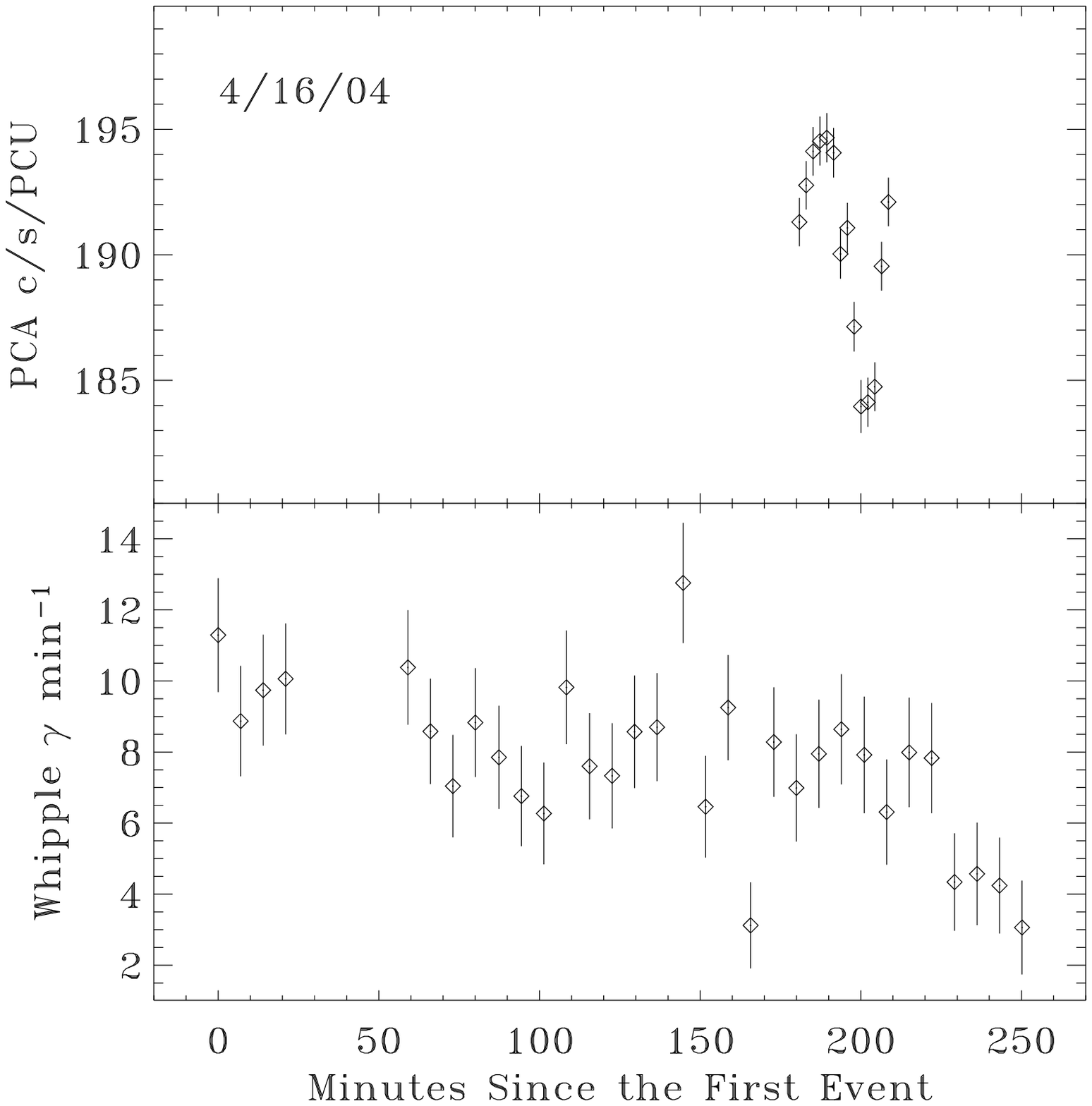,width=3.0in}
\caption{{\em (left)} The most rapid X-ray flare detected during the 
campaign. {\em (right)} Another rapid X-ray flare coincide with the 
occurrence of a longer-duration TeV flare. The gamma-ray light curve 
was made with 7-min bins. Note that neither the rising phase nor the 
decaying phase of the TeV flare is adequately sampled due to 
observational constraints. }
\end{figure}

\begin{figure}
\psfig{figure=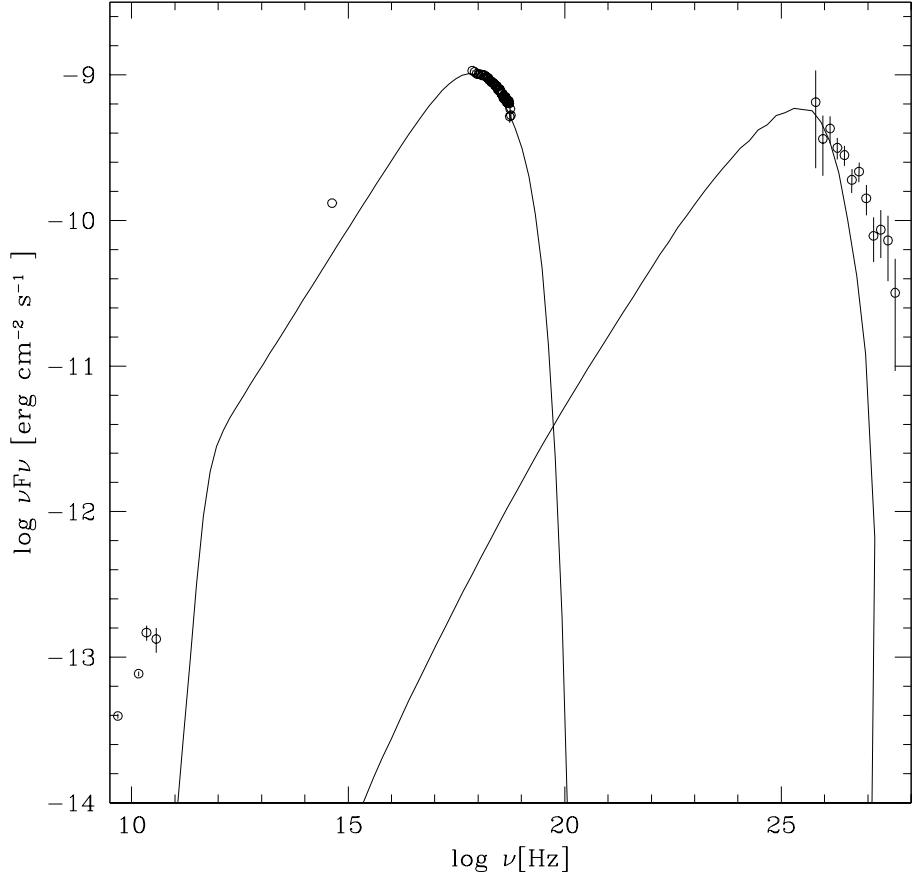,width=5in}
\caption{Spectral energy distribution of Mrk 421. It was derived from the 
high-flux data group (see text). The solid line shows the best-fit to 
the data with a one-zone SSC model (see text): $\delta$=14, $B=0.26$G, 
$R=0.7 \times 10^{16}$cm, $w_e$= 0.086 $ergs\mbox{ }cm^{-3}$,
$p_1$=2.05, $p_2$=3.4, log($E_b$)=11.0, log($E_{min}$)=6.5, and
log($E_{max}$)=11.6. All energies ({\em E's}) are in units of {\em eV}.
Note large deviations at both very low and high frequencies. }
\end{figure}

\begin{figure}
\psfig{figure=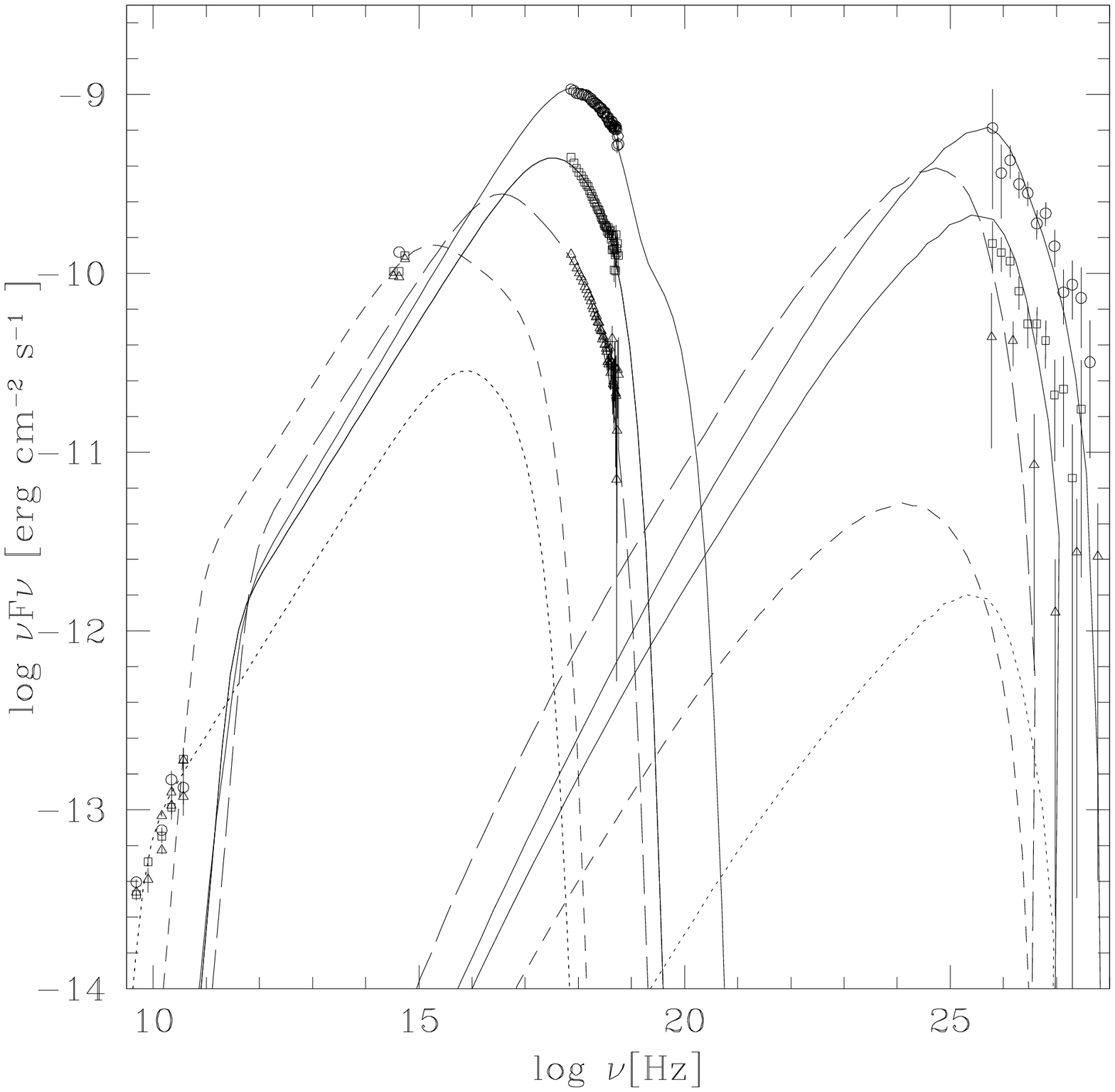,width=5in}
\caption{Spectral energy distributions of Mrk 421 at different X-ray fluxes: 
low (in triangles), medium (in squares), and high (in circles). Fits to the 
SEDs with a multi-zone SSC model (see text) are also shown, in long-dashed 
line for the low-flux group and in solid line for both the medium-flux and 
high-flux groups. 
Parameters used: for the low-flux group, $\delta$=10.0, $B=0.405$ G, 
$R=0.7 \times 10^{16}$cm, $w_e$=0.13777 $ergs\mbox{ }cm^{-3}$, and the 
electron spectral distribution 
(broken power law): $p_1$=2.05, $p_2$=3.6, log($E_b$)=10.34 eV, 
log($E_{min}$)=6.5, and log($E_{max}$)=11.22;
for the medium-flux group: $\delta$=17.8, $B=0.102$G,
$R=1.0 \times 10^{16}$cm, $w_e$=0.03192 $ergs\mbox{ }cm^{-3}$, $p_1$=2.05,
$p_2$=3.4, log($E_b$)=10.98, log($E_{min}$)=6.5, and log($E_{max}$)=11.55;
for the high-flux group: two zones are needed to fit the X-ray and gamma-ray
spectra-- zone1: $\delta$=14, $B=0.046$G, $R=0.7 \times 10^{16}$cm,
$w_e$=0.28 $ergs\mbox{ }cm^{-3}$, $p_1$=2.0, $p_2$=3.0, log($E_b$)=11.5,
log($E_{min}$)=6.5, and log($E_{max}$)=12.5;
zone2: $\delta$=14,
$B=0.11$G, $R=1.5 \times 10^{16}$cm, $w_e$=0.025 $ergs\mbox{ }cm^{-3}$,
$p_1$=2.0, $p_2$=3.0, log($E_b$)=11.0, log($E_{min}$)=6.5, and
log($E_{max}$)=11.6;
plus two additional zones for fitting the radio and optical spectra:
radio (in dotted line): $\delta$=14, $B=0.003$G, $R=8.0\times 10^{17}$cm, 
$w_e$= 0.000012 $ergs\mbox{ }cm^{-3}$, $p_1$=2.05,
$p_2$=3.4, log($E_b$)=11.0, log($E_{min}$)=6.5, and
log($E_{max}$)=11.5;
optical (in dashed line): $\delta$=15,
$B=0.055$G, $R=1.6 \times 10^{17}$cm, $w_e$= 0.00012 $ergs\mbox{ }cm^{-3}$,
$p_1$=2.05, $p_2$=3.4, log($E_b$)=10.0, log($E_{min}$)=6.5, and
log($E_{max}$)=11.0. All energies ({\em E's}) are in units of {\em eV}. }

\end{figure}


\clearpage

\begin{references}

\reference{} Aharonian, F. 2000, New Astronomy, 5, 377
\reference{} Aharonian, F., et al. 2002, A\&A, 393, 89 
\reference{} Aharonian, F., et al. 2003, A\&A, 403, L1
\reference{} Alexander, T., 1997, in Proc. ``Astronomical Time Series'', Eds. D.~Maoz, A.~Sternberg, \& E.~M.~Leibowitz, (Dordrecht: Kluwer), 163 
\reference{} Aller, H. D., Aller, M. F., Latimer, G. E., \& Hodge, P. E. 1985, ApJS, 59, 513 \
\reference{} Baars, J.W.M., Genzel, R, Pauliny-Toth, I.I.K., \& Witzel, A. 1977, A\&A, 61, 99 
\reference{} Beall,~J.~H., \& Bednarek,~W. 1999, ApJ, 510, 188
\reference{} B\"ottcher, M., et al., 1997, A\&A, 324, 395 
\reference{} B\"{o}ttcher,~M. 2002, in Proc. ``The Gamma-ray Universe'', XXII Moriond Astrophysics Meeting, eds. A. Goldwurm et al., p. 151 
\reference{} B\"{o}ttcher,~M. 2005, ApJ, 621, 176 
\reference{} Buckley,~J.~H., et al. 1996, ApJ, 472, L9
\reference{} Catanese,~M., et al. 1997, ApJ, 487, L143
\reference{} Cui,~W. 2004, ApJ, 605, 662 
\reference{} Cui,~W., et al. 2004, in Proc. "International Symposium on High Energy Gamma-Ray Astronomy" (Gamma-2004), eds. F.A. Aharonian and H. Voelk, AIP Conf. Ser., 745, 455 (astro-ph/0410160) 
\reference{} Daniel,~M., et al. 2005, ApJ, 621, 181 
\reference{} Dar,~A., \& Laor,~A. 1997, ApJ, 478, L5
\reference{} Dermer, C. D., et al. 1992, A\&A, 256, L27
\reference{} Dicke,~J, \& Lockman,~J. 1990, ARA\&A. 28, 215 
\reference{} Dondi,~L., \& Ghisellini,~G. 1995, MNRAS, 273, 583 
\reference{} Edelson,~R.~A., \& Krolik, J. H. 1988, ApJ, 333, 646
\reference{} Edelson,~R.~A., et al. 1996, ApJ, 470, 364 
\reference{} Falcone,~A., et al. 2004, ApJ, 613, 710 
\reference{} Finley, J. P., et al. 2001, in Proc. 27th Int. Cosmic Ray Conf., 199 
\reference{} Fossati, G., et al. 1998, MNRAS, 299, 433 
\reference{} Gaidos,~J.~A., et al. 1996, Nature, 383, 319 
\reference{} Giommi, P., Barr, P., Pollock, A. M. T., Garilli, B., \& Maccagni, D. 1990, ApJ, 356, 432 
\reference{} Hillas, A. M., et al. 1998, ApJ, 503, 744 
\reference{} Krawczynski,~H., et al. 2004, ApJ, 601, 151 
\reference{} Krennrich.~F., et al. 2002, ApJ, 575, L9 
\reference{} LeBohec,~S., \& Holder,~J. 2003, Astropart. Phys., 19, 221
\reference{} Landolt, A. 1992, AJ, 104, 340 
\reference{} Lyutikov,~M. 2003, New Astr. Rev. 47, 513 
\reference{} Mannheim,~K., \& Biermann,~P.~L. 1992, A\&A, 253, L21
\reference{} Maraschi, L., Ghisellini, G., \& Celotti, A., 1992, ApJ, 397, L5
\reference{} Maraschi,~L., et al. 1999, Astropart. Phys., 11, 189
\reference{} Marscher, A. P., \& Gear, W. K., 1985, ApJ, 298, 11 
\reference{} Mastichiadis, A., \& Kirk, J. G., 1997, A\&A, 320, 19 
\reference{} MacMinn, D., \& Primack, J. R. 1996, Space Sci. Rev., 75, 413
\reference{} Mochejska, B.~J., Stanek, K.~Z., Sasselov, D.~D., \& Szentgyorgyi, A.~H.\ 2002, AJ, 123, 3460 
\reference{} Mohanty, G., et al. 1998, Astropart. Phys., 9, 15 
\reference{} M\"ucke, A., et al. 2003, Astropart. Phys., 18, 593 
\reference{} Nilsson,~K., Pursimo,~T., Takalo,~L.~O., Sillanp\"a\"a,~A., 
\& Pietil\"a,~H. 1999, PASP, 111, 1223
\reference{} Petry, D., et al., 2000, ApJ, 536, 742
\reference{} Petry, D., et al., 2002, ApJ, 580, 104
\reference{} Pian,~E., et al. 1998, ApJ, 492, L17 
\reference{} Pohl,~M., \& Schlickeiser,~R. 2000, A\&A, 354, 395
\reference{} Punch,~M., et al. 1992, Nature, 358, 477 
\reference{} Rees, M. J., 1978, MNRAS, 184, P61 
\reference{} Reynolds, P., et al. 1993, ApJ, 404, 206 
\reference{} Rybicki,~G.~B., \& Lightman,~A.~P., Radiative Processes in
Astrophysics (New York: John Wiley \& Sons)
\reference{} Sikora, M., et al. 1994, ApJ, 421, 153 
\reference{} Spada, M., et al., 2001, MNRAS, 325, 1559 
\reference{} Stetson, P. B. 1987, PASP, 99, 191 
\reference{} Ter\"{a}sranta,~H., et al. 1998, A\&AS, 132, 305 
\reference{} Urry,~C.~M., \& Padovani,~P. 1995, PASP, 107, 803 \
\reference{} Villata,~M., Raiteri,~C.~M., Lanteri,~L., Sobrito,~G., \& Cavallone,~M. 1998, A\&AS, 130, 305
\reference{} Weekes,~T.~C. 2003, Proc. 28th ICRC (astro-ph/0312179) 
\reference{} Xue,~Y., \& Cui,~W. 2005, ApJ, 622, 160

\end{references}
\end{document}